\documentclass[aps,prl,twocolumn,superscriptaddress,floatfix,citeautoscript,preprintnumbers,longbibliography]{revtex4-1}
\usepackage[dvipsnames]{xcolor}
\usepackage{amsmath,amssymb}
\usepackage{bbold}
\usepackage{physics}

\usepackage[colorlinks = true,
            linkcolor = blue,
            urlcolor  = blue,
            citecolor = blue,
            anchorcolor = blue]{hyperref}
\usepackage{cleveref}
\usepackage{graphicx}
\usepackage{dsfont}
\usepackage{epstopdf}
\begin{document}
\crefname{equation}{Eq.}{Eqs.}
\crefname{figure}{Fig.}{Fig.}
\crefname{appendix}{Appendix}{Appendix}
\newcommand{\widesim}[2][3]{
  \mathrel{\overset{#2}{\scalebox{#1}[1]{$\sim$}}}
}

\title{Efficient Adiabatic Preparation of Tensor Network States}
\author{Zhi-Yuan Wei}
\author{Daniel Malz}
\author{J. Ignacio Cirac}
\affiliation{Max-Planck-Institut f{\"u}r Quantenoptik, Hans-Kopfermann-Stra{\ss}e 1, D-85748 Garching, Germany}
\affiliation{Munich Center for Quantum Science and Technology (MCQST), Schellingstr. 4, D-80799 M{\"u}nchen, Germany}
\date{\today}

\begin{abstract} 
We propose and study a specific adiabatic path to prepare those tensor network states that are unique ground states of few-body parent Hamiltonians in finite lattices, which include normal tensor network states, as well as other relevant non-normal states. This path guarantees a gap for finite systems and allows for efficient numerical simulation. In 1D we numerically investigate the preparation of a family of states with varying correlation lengths and the 1D AKLT state and show that adiabatic preparation can be much faster than standard methods based on sequential preparation. We also apply the method to the 2D AKLT state on the hexagonal lattice for which no method based on sequential preparation is known, and show that it can be prepared very efficiently for relatively large lattices.
\end{abstract}

\maketitle

Matrix Product States (MPS)~\cite{fannes1992finitely,10.5555/2011832.2011833}, and more generally, Projected Entangled-Pair States (PEPS)~\cite{verstraete2004renormalization}, capture the physical properties of systems obeying the entanglement area law~\cite{eisert2010colloquium}. PEPS contain a rich set of many-body states~\cite{cirac2021matrix} such as the cluster state~\cite{Briegel2001}, toric codes~\cite{Kitaev2003}, GHZ state~\cite{greenberger1989going} and W state~\cite{dur2000three} in quantum information, or the AKLT states~\cite{Affleck1987, Affleck1988}, valence-bond states~\cite{anderson1987resonating} and string net states~\cite{levin2005string} in condensed matter physics. There is thus increasing interest in finding ways of preparing them in quantum computers or quantum simulators, either for quantum information applications like computing ~\cite{briegel2009measurement}, metrology~\cite{Jarzyna2013}, communication and networking~\cite{Azuma2015}, or as variational states for the study of many-body quantum systems~\cite{Huggins2019}.

MPS are most naturally prepared sequentially~\cite{Schon2005}, which requires a time that scales linearly in the number of sites $N$. In higher dimensions, for PEPS, this is not possible in general~\cite{Schuch2007}. However, certain subclasses of PEPS can be generated sequentially in linear time~\cite{Banuls2008,Pichler2017,Zaletel2020,zypp}. 
Sequential preparation has been used in various platforms to experimentally prepare MPS and PEPS~\cite{schwartz2016,Besse2020,satzinger2021,smith2022crossing}.

Besides quantum circuits, adiabatic algorithms are also widely used to prepare many-body states on quantum devices~\cite{Albash2018}. By smoothly tuning the Hamiltonians along a gapped path that connects a trivial state to the target state, quasi-adiabatic evolution for a time $T$ produces a state very close to the target state. Adiabatic algorithms have been proposed to prepare PEPS~\cite{Schwarz2012a,Schwarz2013,Ge2016,cruz2022preparation}, and in particular, Ref.~\cite{Ge2016} proved that it is possible to prepare a generic family of them, so-called \textit{normal} PEPS~\cite{fannes1992finitely,10.5555/2011832.2011833}, in time $T = O\left( {{\rm polylog} N} \right)$ with a specific method that switches on and off certain Hamiltonian terms adiabatically and provided there exists a gap along the whole path that is lower bounded by a constant. A method to compute such a lower bound based on semidefinite programming has been presented in Ref.~\cite{cruz2022preparation}. While those methods provide rigorous proofs for the asymptotic limit $N\gg 1$, it is not clear how they perform in practice, in particular for the intermediate sizes available in the near term. For such cases, there is no guarantee that they provide any advantage with respect to sequential methods.

In this paper, we propose a specific adiabatic path to prepare PEPS in any dimension that are unique ground states of local frustration-free Hamiltonians and analyze its performance. Our path guarantees the existence of a gap (for finite systems), and in contrast to Ref.~\cite{Ge2016,cruz2022preparation} yields Hamiltonian with substantially smaller support for preparing the 2D AKLT state on the hexagonal lattice. Moreover, it extends to certain non-normal PEPS~\cite{supp}, which allows us to prepare the AKLT states in arbitrary geometries.

Since the ground states along our path are always PEPS, we are able to simulate relatively large systems, which we use to numerically determine the performance of the algorithm. In 1D, we consider the family of MPS introduced in Ref.~\cite{Wolf2006}, which allows us to investigate how the efficiency of the algorithm depends on correlation length. We also consider the paradigmatic 1D AKLT state. We obtain that for system sizes up to $N=5000$, the preparation can be much more efficient than sequential preparation, with $T \sim \textrm{polylog } N$ in the regime we study~\footnote{Note that an asymptotic scaling of this kind is not possible with the adiabatic ramps we use in this paper, as shown in~\cite{nenciu1993linear,hagedorn2002elementary,Rezakhani2010} and the appendix~\cite{supp}.}. In 2D, our adiabatic path overcomes several difficulties and allows us to simulate the adiabatic preparation of the 2D AKLT state on the hexagonal lattice up to $N\sim 10\times 10$. Our results indicate that adiabatic preparation is very efficient also in higher dimensions.

\begin{figure}[h!]
	\centering
	\includegraphics[width=0.5\textwidth]{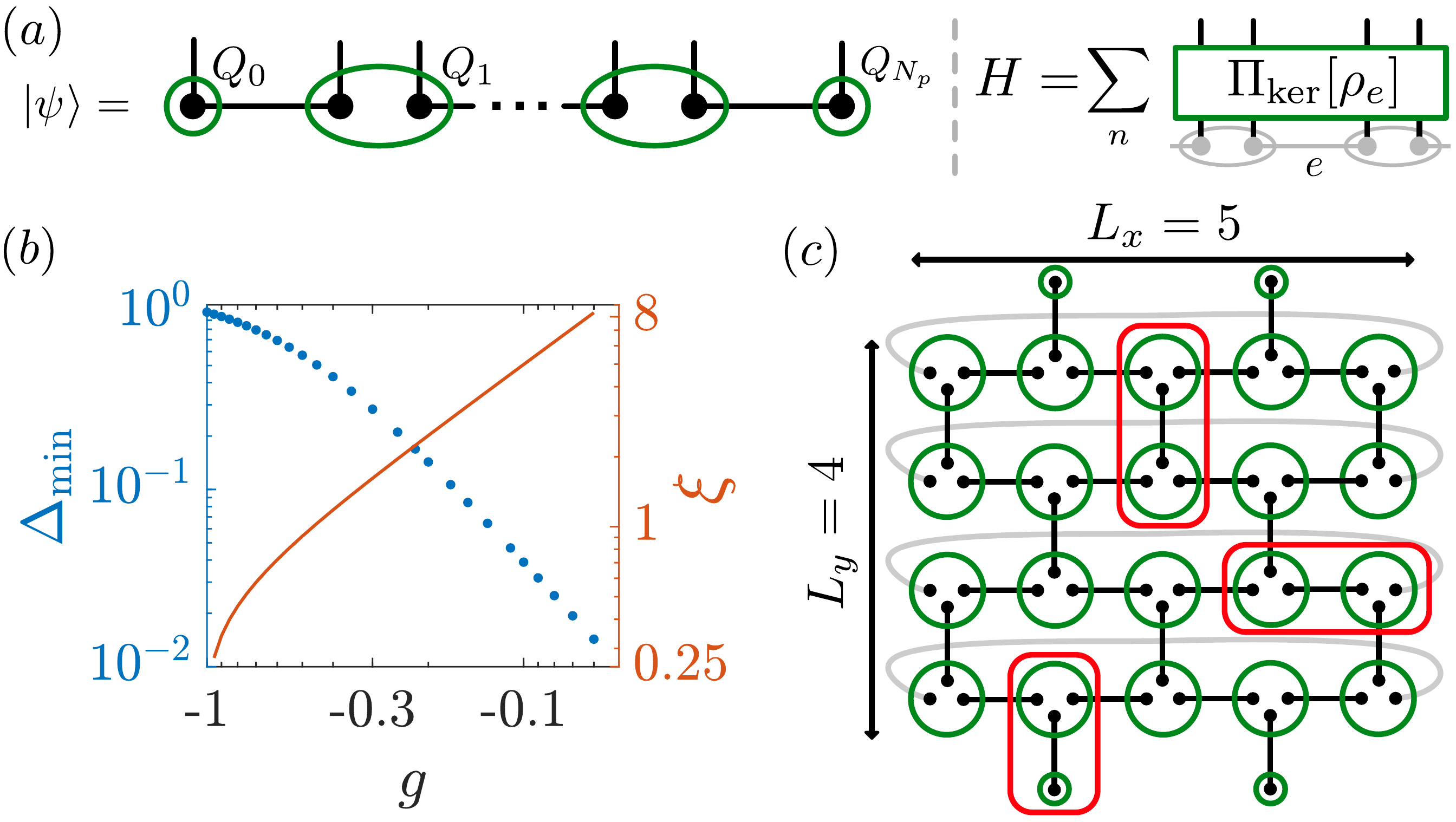}
        \caption{Adiabatic preparation of MPS and PEPS. (a) In 1D, a MPS $|\psi\rangle$ [\cref{psi_f_targ}] is constructed by applying a set of operators $\{Q_v\}$ (green circles) on a product of pairs of maximally entangled virtual qudits (the connected dots). $|\psi\rangle$ is the ground state of a local Hamiltonian $H$ [\cref{tns_ham}], which can be taken to be the sum of projectors (green rectangle) onto the kernel of the corresponding reduced density matrices.
(b) The minimal gap $\Delta _ {\rm min}$ of the adiabatic path [c.f.~\cref{tns_q}] (computed with $N=400$ sites here, but it is size-independent~\cite{supp}) and the correlation length $\xi$ for states in the MPS family [c.f.\cref{cls_mps}]. (c) 2D PEPS on the hexagonal lattice. The green circles denote the operators $\{Q_v\}$, and the connected dots denote maximally entangled virtual qudit pairs. Each term of the parent Hamiltonian $H$ acts on neighboring sites (shown as red rectangles). The size of the lattice is $L_x \times L_y$. In our numerics, we focus on the 2D AKLT state on the hexagonal lattice with cylinder boundary conditions (illustrated through gray lines).}
        \label{fig1}
\end{figure}

\textit{PEPS.---}
PEPS can be built by applying local commuting operators to a product state of maximally entangled pairs in a lattice~\cite{verstraete2004renormalization,Molnar2018}. Let us consider a regular lattice denoted by a graph $\cal G$, with edges $\cal E$ and sites $\cal V$. The coordination number of site $v\in {\cal V}$ is $n_v$, i.e. each site contains $n_v$ virtual qudits. Defining local operators $\{Q_v\}$ that map the $D$-level virtual qudits on site $v\in {\cal V}$ to a $d$-level physical site, the PEPS is expressed as [see \cref{fig1}(a) for 1D case and \cref{fig1}(c) for 2D hexagonal lattice case]
\begin{equation} \label{psi_f_targ}
	\left|\psi\right\rangle \propto{ \bigotimes_{v \in \cal V} } Q_{v} {\bigotimes_{e \in \cal E} \left| {{\Phi ^ +  }} \right\rangle _e },
\end{equation}
with $\left|\Phi^{+}\right\rangle \propto \sum_{\alpha =0}^{D-1} |\alpha \alpha\rangle$. Here $D$ is the bond dimension of the PEPS, and $d$ is the physical dimension. For instance, MPS can be viewed as 1D PEPS with $n_v=2$ virtual qudits per site [c.f.~\cref{fig1}(a)].  The matrix representation of $\{Q_v\}$ in the bulk for MPS then reads
\begin{equation} \label{proj_qk}
Q_v^{\rm 1D} = \sum\limits_{i_v = 0}^{d-1} {\sum\limits_{\alpha_v ,\beta_v = 0}^{D-1} {A_{[v] \alpha_v \beta_v }^{i_v}\left| i_v \right\rangle \left\langle {\alpha_{v} \beta_{v} } \right|} }.
\end{equation}
The operators on the boundary each act on a single qudit~\footnote{For all states we study in this paper, we simply set $\{Q_v\}$ on the boundary to be identity operators.}. By blocking neighboring sites, we can enlarge the physical dimension such that $d \geq D^{n_v}$. In this case, without loss of generality, we can apply a polar decomposition to write $\{Q_v\}$ as positive-semidefinite operators with $d=D^{n_v}$, which holds for arbitrary PEPS up to a layer of local isometries~\cite{supp}. A PEPS is called \textit{injective} if $\{Q_v\}$ are left-invertible~\cite{fannes1992finitely,10.5555/2011832.2011833}. If the operators obtained after blocking a finite number of sites are invertible, the PEPS is called \textit{normal}.

In this paper, we aim to prepare a large class of PEPS that are unique ground states of local frustration-free Hamiltonians. This includes all normal (and thus all injective) PEPS~\cite{fannes1992finitely,10.5555/2011832.2011833}, but also other relevant states like the AKLT states (possibly non-normal~\cite{Molnar2018,supp}), where a much simpler parent Hamiltonian is known~\cite{Affleck1987, Affleck1988}. In particular, we consider the following parent Hamiltonian~\cite{Wolf2006,perez2008peps} [c.f.~\cref{fig1}(a)]
\begin{equation} \label{tns_ham}
H = \sum_{e} {\Pi}_{\rm ker}[\rho_{e}],
\end{equation}
where ${\Pi}_{\rm ker}$ projects on the kernel of $\rho_{e}$, which is the reduced density matrix of neighboring sites around the edge $e \in {\cal E}$~\footnote{The Hamiltonian [\cref{tns_ham}] can be efficiently constructed for arbitrary 1D MPS and 2D translationally invariant (TI) PEPS, which covers all examples studied in this paper. However, for a generic non-TI 2D PEPS, one cannot scalably obtain the few-sites reduced density matrices $\{\rho_e\}$. In this case, it is possible to use a numerical optimization approach~\cite{Giudici2022rkq} to obtain a parent Hamiltonian with each term also only support on a few sites.}. Note that $\Vert {\Pi}_{\rm ker}[\rho_{e}] \Vert = 1$, thus the time is unit-less in this paper.

The parent Hamiltonian $H$ [\cref{tns_ham}] for injective PEPS has a unique ground state~\cite{perez2008peps}, which implies a nonzero gap that may depend on the system size $N$. Moreover, $H$ for 1D injective MPS is guaranteed to be gapped also in the thermodynamic limit~\cite{fannes1992finitely}. Finally, $H$ for the AKLT states is equivalent to the known two-body parent Hamiltonian~\cite{Affleck1987, Affleck1988}.

\textit{Examples.---}
We study two paradigm examples of PEPS in this paper. The first example is a family of MPS of bond dimension $D=2$~\cite{Wolf2006}. In this case the graph $\cal G$ corresponds to a chain formed by $N=2N_p$ qubits forming $N_p$ pairs [c.f.~\cref{fig1}(a)]. After blocking each neighboring two sites, we arrive at the injective form of the MPS family for $g\neq 0$ (with $d=D^2=4$), where the matrices in \cref{proj_qk} are given through
\begin{align} \label{cls_mps}
	A_{[v]}^{0}(g) &=\left(\begin{array}{cc}
0 & 0\\
1 & 1
\end{array}\right),\quad A_{[v]}^{1}(g)=\left(\begin{array}{cc}
0 & 0\\
1 & g
\end{array}\right), \\
 A_{[v]}^{2}(g)&=\left(\begin{array}{cc}
g & g\\
0 & 0
\end{array}\right),\quad A_{[v]}^{3}(g)=\left(\begin{array}{cc}
1 & g\\
0 & 0
\end{array}\right).\nonumber
\end{align}
The corresponding parent Hamiltonian [\cref{tns_ham}] acts only on nearest neighbors, but with each site containing two qubits.

We will study the preparation of states with $g \in (-1,0)$, which interpolates between the cluster state ($g=-1$) and the GHZ state ($g=0$). For $g<0$, the correlation length $\xi$ of the MPS family can be obtained as [c.f.~\cref{fig1}(b)]
\begin{equation} \label{qpt_cor_g}
\xi  = \left( {\ln  {{\frac{{ 1 - g}}{{1 + g}}} } } \right)^{-1}.
\end{equation}
Thus by tuning $g$, we can explore the effect of correlation length on the performance of the adiabatic algorithm. Note that $g \in (-1,0)$ already covers all states with $g<0$, since the tensors $\{ A(g)\}$ in \cref{cls_mps} can be mapped to $\{ A(1/g)\}$ by a gauge transformation~\footnote{One can transform $\{A_{g}\}$ to $\{A_{1/g}\}$ by first swapping the two indices in the auxiliary space (corresponding to transposing the matrices $\{A\}$), then swapping the physical index 2 and 3, and finally normalize the state.}.

The other example we consider is the 1D AKLT state of spin $S=1$ and the 2D AKLT state of spin $S=3/2$ in the hexagonal lattice [c.f.~\cref{fig1}(c)]~\cite{Affleck1987, Affleck1988}. AKLT states can be formed by first having a product state of singlets consisting of virtual qubits that connect neighboring sites of the lattice, then projecting the virtual qubits at each site $v$ to their symmetric subspace. AKLT states can be written as $D=2$ PEPS [\cref{psi_f_targ}], and we promote the virtual qubits into physical ones, such that the operators $\{Q_v\}$ are already positive-semidefinite without blocking~\cite{supp}.

\textit{Adiabatic algorithm.---} We propose an adiabatic path parametrized by $s$, which connects a product state of maximally entangled pairs $\left|\psi(0) \right\rangle \equiv {\bigotimes_{e \in \cal E} \left| {{\Phi ^ +  }} \right\rangle _e }$~\footnote{The initial state $\left|\psi(0) \right\rangle$ can be prepared with local quantum circuit or adiabatic evolution of a constant depth (time).} to the target PEPS $\left|\psi(1) \right\rangle \equiv \left|\psi\right\rangle$. We choose the instantaneous ground states $\left|\psi(s)\right\rangle$ in this path to be always PEPS of bond dimension $D$, with [see \cref{psi_f_targ}]
\begin{equation} \label{tns_q}
	Q_v(s)=s  Q_{v}+(1-s)  {\mathbb 1}.
\end{equation}
For all $s$, one can construct the parent Hamiltonian $H(s)$ [c.f.~\cref{tns_ham}] such that $\left|\psi(s)\right\rangle$ is its ground state.
This path has the following features:
\begin{enumerate}
	\item Adiabatic evolution along this path can be classically simulated (approximately), as its instantaneous ground states $\left|\psi(s)\right\rangle$ [c.f.~\cref{tns_q}] are PEPS of bond dimension $D$~\cite{verstraete2004renormalization}.
	\item This path is gapped for finite systems. First, for $s\in [0,1)$, $\{Q_v(s)\}$ are invertible (since $\{Q_v\}$ are positive-semidefinite), thus the Hamiltonian $H(s)$ along the path has a non-zero gap $\Delta(s)>0$. For $s=1$ we also have $\Delta(1)>0$ since we consider the class of PEPS that are unique ground states of local Hamiltonians [\cref{tns_ham}].	As for finite systems, $\Delta(s)$ is continuous and differentiable in the whole interval $s\in [0,1]$ and the derivative $d \Delta(s)/ds$ is finite, it immediately implies that $\Delta(s) \ge  \Delta_{\rm min} >0$ (note that $\Delta_{\rm min}$ may depend on the system size $N$).
	\item The support of each term in the Hamiltonian $H(s)$ [\cref{tns_ham}] stays the same for all $s \in [0, 1]$, which may simplify the experimental implementation. For example, for preparing AKLT states, $H(s)$ is always two-body.
\end{enumerate}

In the following, we study the adiabatic preparation of these examples using our path.

\textit{Preparation of the MPS family.---} First, we computed the minimal gap $\Delta_{\rm min}$ during the adiabatic path [c.f.~\cref{tns_q}] for the MPS family [c.f.~\cref{cls_mps}] in \cref{fig1}(b). One can see that $\Delta_{\rm min}$ decreases as the correlation length $\xi$ increases, which suggests that the adiabatic algorithm should perform better when the correlation length is smaller. 

We classically simulate the quasi-adiabatic time evolution of a chain of $N$ qubits for a time $T$~\cite{supp}, following the path in \cref{tns_q}, where we take the interpolation function $s(t/T)_{1\rm D} \equiv {\sin ^2}\left[ \pi/2 \cdot {\sin ^2} (\pi t/{2T})  \right]$~\footnote{We choose the interpolation functions such that they have a finite level of smoothness, which is both experimentally realistic and lead to an almost exponential decay of the error [c.f.~\cref{cls_qpt_fid_scale}] in a wide range of evolution time $T$. Since in we need to reach larger system size $N\sim 5000$ in 1D compared to $N\sim 100$ in 2D, we choose the function $s(t,T)_{\rm 1D}$ to be more smooth than $s(t,T)_{\rm 2D}$ at $t=0$ and $t=T$.}.

Assuming that the state we obtained after the evolution is $| {{ \phi (T)}} \rangle$, its fidelity ${\cal F}$ compared to the target state $\left| {{\psi(1)}} \right\rangle$ is ${| {\langle \psi(1) | {{\phi (T)}} \rangle } |^2}$. For all fixed $T\geq 0$, we numerically find that ${\cal F}$ decays exponentially with system size $N$~\cite{supp}, which allows us to define an error density $\kappa(T)$ that is independent of system size. Thus we can write
\begin{equation} \label{fid_n_scale}
{\cal F}(N,T)=\exp\left[-\kappa(T)\cdot N-c(T)\right],
\end{equation}
where $c(T)$ is an error that comes from the boundaries of the system and is independent of $N$. This indicates that during the adiabatic dynamics, the errors in different regions of the chain change almost uniformly.

\label{qpt_ak_sec}
\begin{figure}[h!]
	\centering
	\includegraphics[width=0.5\textwidth]{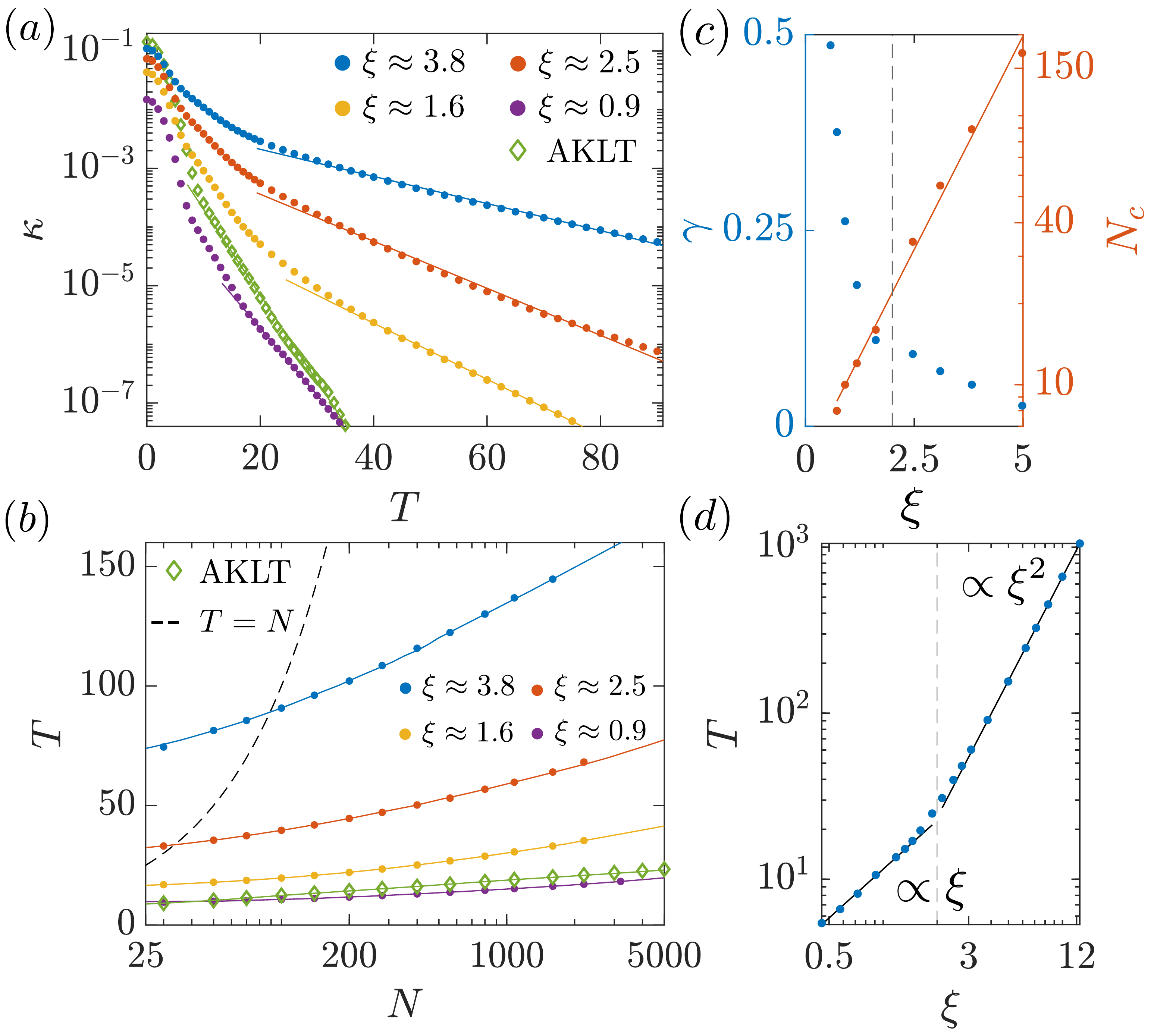}
        \caption{(a) The error density $\kappa(T)$ [c.f.~\cref{fid_n_scale}] as a function of the adiabatic evolution time $T$, for preparing states in the MPS family with various correlation length $\xi$ as well as the 1D AKLT state. The lines are exponential fits to the data.
(b) Time $T$ for preparing the same set of states in (a) of size $N$ with fidelity ${\cal F}=0.99$. The dots are obtained by the numerical simulation, while the solid lines are the prediction of \cref{fid_n_scale}. The dashed line denotes the scaling for the sequential method (assuming it takes a time $T=N$).
(c) The decay rate $\gamma$ [c.f.~\cref{cls_qpt_fid_scale}] and the system size $N_c$ where the adiabatic method and sequential method give the same preparation time $(T=N_c)$ as a function of the correlation length $\xi$ for the MPS family. The vertical dashed lines in (c,d) correspond to the size of each lattice site ($\xi=2$). (d) Time $T$ to prepare states in the MPS family of fixed particle number $N=100$, as a function of the correlation length $\xi$.}
        \label{fig2}
\end{figure}
The error density $\kappa(T)$ can be obtained by fitting the fidelity of preparing the same state of various system sizes $N$ and a fixed time $T$ using the scaling \cref{fid_n_scale}~\cite{supp}. In \cref{fig2}(a) we show $\kappa(T)$ for the MPS family as a function of $T$, and it features two regimes. When $T$ is small, the dynamics is not adiabatic, and we see $\kappa(T)$ already starts to decay quickly. When $T$ becomes larger, $\kappa(T)$ enters a regime of almost exponential decay, which we fit with
\begin{equation} \label{cls_qpt_fid_scale}
{\kappa(T)} \approx \kappa_0 \exp (- \gamma T).
\end{equation}
The decay rate $\gamma$ decreases with increasing correlation length $\xi$ [see \cref{fig2}(c)], and the boundary term $|c(T)|$ shows a similar behavior as $\kappa(T)$~\cite{supp}. Note that, due to the finite smoothness of the interpolation function $s(t/T)_{\rm 1D}$ we use, we expect that $\kappa \sim 1/{\rm poly}(T)$ when $T\rightarrow \infty$~\cite{nenciu1993linear,hagedorn2002elementary,lidar2009,Rezakhani2010,supp}. One can extend the range of the exponential decay by making the interpolation function smoother, however at the expense of reducing the decay rate $\gamma$~\cite{supp}.

In \cref{fig2}(b), we show the dependence of time $T$ required to prepare the given target state with fidelity ${\cal F}=0.99$ on system size $N$ up to $N \sim 5000$. The results agree well with the simple expression \cref{fid_n_scale}, which lead to $T \sim \textrm{polylog } N$ in this regime (which is the relevant regime experimentally).
We also compare the adiabatic preparation to sequential preparation~\cite{Schon2005}, which we assume takes a time $T=N$ [see \cref{fig2}(b)]. One sees that the adiabatic algorithm outperforms the sequential preparation method in terms of the preparation time $T$ when the system size $N$ is larger than a threshold value $N_c(\xi)$~\footnote{Note that the sequential method prepares the state deterministically, while the adiabatic method always prepares the state approximately (with high fidelity), which is enough for practical purposes. Moreover, there could be a constant overhead to implement the Hamiltonian dynamics using quantum circuits, and this does not affect our statement that there will be a regime $(N\gg 1)$ where the adiabatic preparation will be more efficient.}. We find numerically that [c.f.~\cref{fig2}(c)] $N_c$ almost grows exponentially with $\xi$, which indicates that when the correlation length is smaller than a fraction of the system size, the adiabatic algorithm prepares the MPS family [c.f.~\cref{cls_mps}] faster than the sequential method. 

Finally, we show the time $T$ to prepare states of system size $N=100$ as a function of the correlation length $\xi$ in \cref{fig2}(d). We observe $T\propto \xi$ ($T\propto \xi^2$) when $\xi$ is smaller (larger) than the length of the lattice site, which is $\xi=2$ since each site contains two qubits. This shows that the size of each lattice site sets another length scale in the system, and one can also see such behavior for the decay rate $\gamma$ shown \cref{fig2}(b).

\textit{Preparation of AKLT states.---} Now we study the preparation of the 1D and 2D AKLT states using our adiabatic path. In 1D, it has been proposed to prepare the AKLT state sequentially~\cite{Schon2005}, dissipatively~\cite{Zhou2021b}, using measurements~\cite{Kaltenbaek2010}, or by parallelly fusing multiple AKLT chains~\cite{Zhou2021b}, which has the best known preparation time $T=O(\log ^2 N)$. In \cref{fig2}(a,b) we show the results for adiabatic preparation of the 1D AKLT state using our adiabatic path. As expected, $T \sim \textrm{polylog } N$ up to $N=5000$.

Preparation of the 2D AKLT states are much less explored. For the case of hexagonal lattice, this state have a gapped parent Hamiltonian~\cite{Lemm2020,Pomata2020}, and there is indirect evidence suggesting that this state can be adiabatically prepared with $T= O(N)$~\cite{koch2015affleck}. The general protocol~\cite{Ge2016,cruz2022preparation} predicts the preparation time $T = O\left( {{\rm polylog} N} \right)$ when $N \gg 1$, but it faces the following challenges: First, the construction of the parent Hamiltonian there requires the target PEPS to be normal, which does not work for non-normal PEPS such as the 2D AKLT state on the square lattice~\cite{Molnar2018}, or leads to a Hamiltonian for the 2D AKLT state on the hexagonal lattice that acts on large clusters~\footnote{The 2D AKLT state studied here requires to block 6 sites in a hexagon [c.f.~\cref{fig1}(c)] to make the tensor injective.}, making it difficult to implement in current devices. More importantly, it is difficult to simulate an adiabatic evolution in 2D since the time cost of classical simulation algorithms typically has heavy dependence on the bond dimension of the underlying PEPS~\cite{Lubasch2014}.

Our adiabatic path overcomes the above problems (partially because we promote the virtual qubits to physical ones~\cite{supp}). The Hamiltonian [c.f.~\cref{tns_ham}] along the whole path is two-body and gapped (note that each site contain 3 qubits). Moreover, the instantaneous ground state [c.f.~\cref{tns_q}] is always a PEPS of bond dimension $D=2$~\footnote{This is also true for the path in Ref.~\cite{Ge2016,cruz2022preparation}. However, the large support of each Hamiltonian term makes it still difficult to simulate the adiabatic dynamics along that path.}.

We classically simulate the preparation of the 2D AKLT state on the hexagonal lattice with cylinder boundary condition, and $N\equiv L_x \times L_y$ sites [c.f.~\cref{fig1}(c)]~\cite{supp} using the interpolation function [c.f.~\cref{tns_q}] $s(t,T)_{\rm 2D}\equiv {\sin}^2(\pi t/2T)$~\footnotemark[6]. 

\begin{figure}[h!]
	\centering
\includegraphics[width=0.4\textwidth]{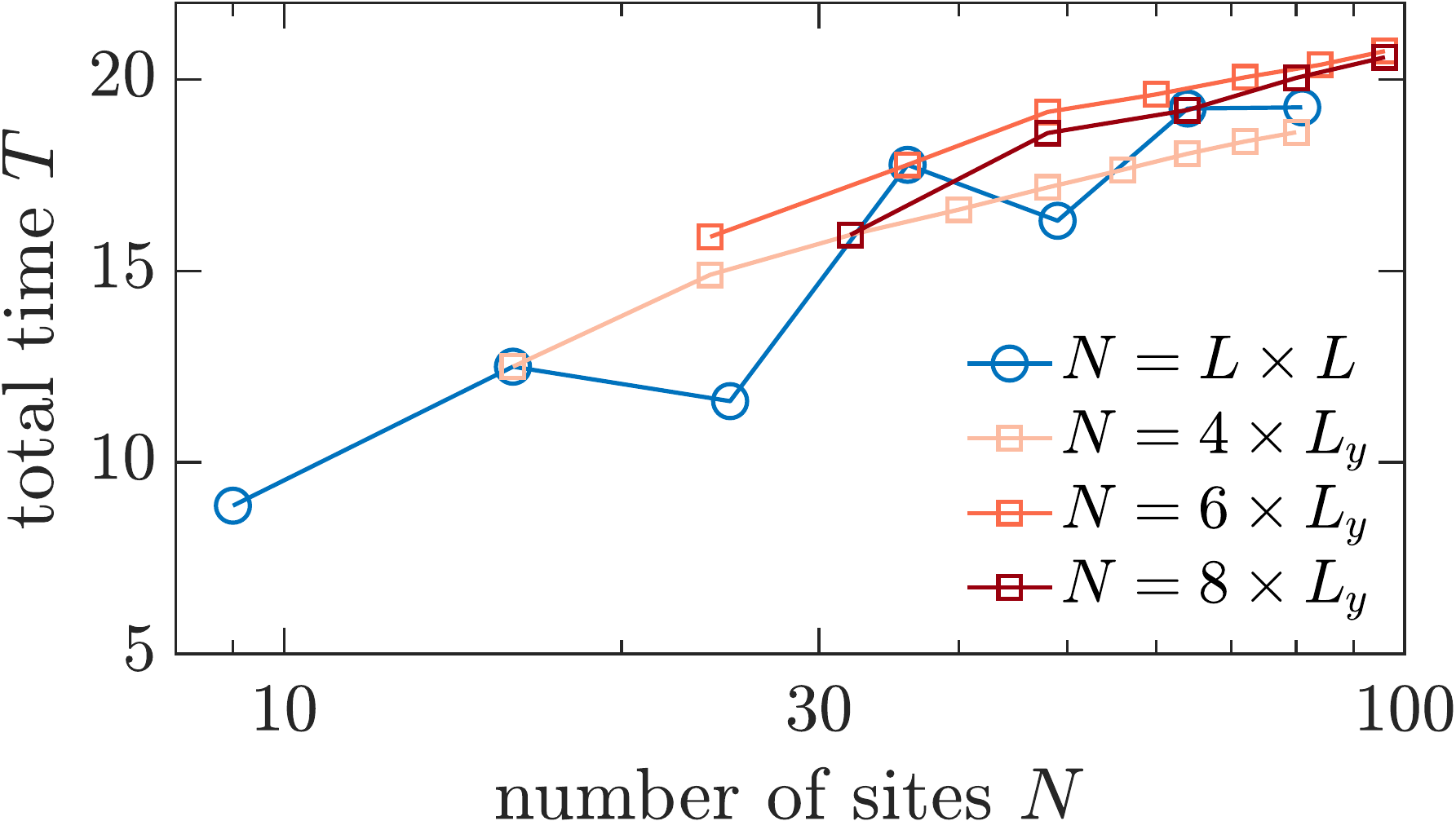}
        \caption{The evolution time $T$ needed to prepare the 2D AKLT state on the hexagonal lattice with fidelity ${\cal F}=0.99$ for $N=L\times L$ ($L=3-9$) and $N=L_x\times L_y$ with various fixed $L_x=4,6,8$. The lines are visual guides.}
        \label{fig3}
\end{figure}

In \cref{fig3} we show the preparation time $T$ needed to reach a fidelity ${\cal F}=0.99$ for different system sizes $N$. In the case of $N=L\times L$, since each hexagon is of size $2\times 1$ [c.f.~\cref{fig1}(c)], for even or odd $L$ the boundary affects $T$ differently. We see that overall $T$ is practically short $(T\sim 10)$, and increases only mildly with system size $N$. We also show $T$ for preparing this state of lattice size $N=L_x \times L$ by fixing different $L_x = 4,6,8$, and observe similar behavior. In particular, the lattice geometry does not strongly affect the preparation time $T$, as it takes a similar $T$ to prepare the state on a $4\times 20$ lattice or $9\times 9 $ lattice.

We expect that $T$ will still be practically short when we increase the system size further than that shown in \cref{fig3}. Moreover, the general nature of the proposed path suggest that this method may be used to efficiently prepare a large variety of (high-dimensional) PEPS. For example, in the SI~\cite{supp} we provide numerical evidence that the 2D AKLT state on the square lattice can be prepared with this method.

\textit{Outlook.---} We have proposed and studied a specific adiabatic path to prepare a large family of MPS and PEPS, which applies to the normal PEPS and other relevant PEPS like the AKLT states.

It is worth checking if the 2D AKLT state of $N \gg 100$ can still be efficiently prepared using quantum devices, and exploring the performance of this adiabatic path to prepare other (potentially non-normal~\cite{Molnar2018}) PEPS. By studying the gap during the adiabatic path in the thermodynamic limit~\cite{Kastoryano2018, cruz2022preparation}, it is possible to probe the asymptotic behavior of the adiabatic algorithm and further improve its performance with adiabatic ramp rate that adapts to the magnitude of the gap. Moreover, in Ref.~\cite{Ge2016} it is shown that adiabatic preparation can also be implemented efficiently on digital quantum computers, which also applies to our results. It is also important to design efficient physical realization of the proposed adiabatic path, which require engineering few-body Hamiltonians. Hamiltonain engineering can be realized in various platforms like the superconducting qubits~\cite{mezzacapo2014many}, ion traps~\cite{bermudez2009competing} and Rydberg atomic arrays~\cite{scholl2022microwave}, which potentially allows to realize large classes of Hamiltonians~\cite{Cubitt2017,Zhou2021a,Choi2020}. Finally, one can study the effect of noise on the adiabatic state preparation. In the presence of noise, we expect the adiabatic method provides an even bigger advantage over the sequential methods for preparing short-range correlated states, since more error accumulates during the sequential preparation (which takes a longer time).

\textit{Acknowledgments.---}
We thank Norbert Schuch and Yilun Yang for their insightful discussions. The research is part of the Munich Quantum Valley, which is supported by the Bavarian state government with funds from the Hightech Agenda Bayern Plus. We acknowledge funding from the German Federal Ministry of Education and Research (BMBF) through EQUAHUMO (Grant No. 13N16066) within the funding program quantum technologies - from basic research to market, and the European Union's Horizon 2020 research and innovation program under Grant No. 899354 (FET Open SuperQuLAN). The numerical calculations were performed using the ITensor Library~\cite{itensor}.

\textit{Note added.---} During completing of this manuscript we became aware of a related protocol that probabilistically creates the 1D and 2D AKLT states using const-depth circuits and post-selection, with a success rate that exponentially decays with the system size $N$~\cite{Murta2022}.

\bibliography{library.bib}

%merlin.mbs apsrev4-1.bst 2010-07-25 4.21a (PWD, AO, DPC) hacked
%Control: key (0)
%Control: author (0) dotless jnrlst
%Control: editor formatted (1) identically to author
%Control: production of article title (0) allowed
%Control: page (1) range
%Control: year (0) verbatim
%Control: production of eprint (0) enabled
\begin{thebibliography}{65}%
\makeatletter
\providecommand \@ifxundefined [1]{%
 \@ifx{#1\undefined}
}%
\providecommand \@ifnum [1]{%
 \ifnum #1\expandafter \@firstoftwo
 \else \expandafter \@secondoftwo
 \fi
}%
\providecommand \@ifx [1]{%
 \ifx #1\expandafter \@firstoftwo
 \else \expandafter \@secondoftwo
 \fi
}%
\providecommand \natexlab [1]{#1}%
\providecommand \enquote  [1]{``#1''}%
\providecommand \bibnamefont  [1]{#1}%
\providecommand \bibfnamefont [1]{#1}%
\providecommand \citenamefont [1]{#1}%
\providecommand \href@noop [0]{\@secondoftwo}%
\providecommand \href [0]{\begingroup \@sanitize@url \@href}%
\providecommand \@href[1]{\@@startlink{#1}\@@href}%
\providecommand \@@href[1]{\endgroup#1\@@endlink}%
\providecommand \@sanitize@url [0]{\catcode `\\12\catcode `\$12\catcode
  `\&12\catcode `\#12\catcode `\^12\catcode `\_12\catcode `\%12\relax}%
\providecommand \@@startlink[1]{}%
\providecommand \@@endlink[0]{}%
\providecommand \url  [0]{\begingroup\@sanitize@url \@url }%
\providecommand \@url [1]{\endgroup\@href {#1}{\urlprefix }}%
\providecommand \urlprefix  [0]{URL }%
\providecommand \Eprint [0]{\href }%
\providecommand \doibase [0]{http://dx.doi.org/}%
\providecommand \selectlanguage [0]{\@gobble}%
\providecommand \bibinfo  [0]{\@secondoftwo}%
\providecommand \bibfield  [0]{\@secondoftwo}%
\providecommand \translation [1]{[#1]}%
\providecommand \BibitemOpen [0]{}%
\providecommand \bibitemStop [0]{}%
\providecommand \bibitemNoStop [0]{.\EOS\space}%
\providecommand \EOS [0]{\spacefactor3000\relax}%
\providecommand \BibitemShut  [1]{\csname bibitem#1\endcsname}%
\let\auto@bib@innerbib\@empty
%</preamble>
\bibitem [{\citenamefont {Fannes}\ \emph {et~al.}(1992)\citenamefont {Fannes},
  \citenamefont {Nachtergaele},\ and\ \citenamefont
  {Werner}}]{fannes1992finitely}%
  \BibitemOpen
  \bibfield  {author} {\bibinfo {author} {\bibfnamefont {Mark}\ \bibnamefont
  {Fannes}}, \bibinfo {author} {\bibfnamefont {Bruno}\ \bibnamefont
  {Nachtergaele}}, \ and\ \bibinfo {author} {\bibfnamefont {Reinhard~F}\
  \bibnamefont {Werner}},\ }\bibfield  {title} {\enquote {\bibinfo {title}
  {{Finitely correlated states on quantum spin chains}},}\ }\href
  {https://link.springer.com/article/10.1007/BF02099178} {\bibfield  {journal}
  {\bibinfo  {journal} {Communications in mathematical physics}\ }\textbf
  {\bibinfo {volume} {144}},\ \bibinfo {pages} {443--490} (\bibinfo {year}
  {1992})}\BibitemShut {NoStop}%
\bibitem [{\citenamefont {Perez-Garcia}\ \emph {et~al.}(2007)\citenamefont
  {Perez-Garcia}, \citenamefont {Verstraete}, \citenamefont {Wolf},\ and\
  \citenamefont {Cirac}}]{10.5555/2011832.2011833}%
  \BibitemOpen
  \bibfield  {author} {\bibinfo {author} {\bibfnamefont {D}~\bibnamefont
  {Perez-Garcia}}, \bibinfo {author} {\bibfnamefont {F}~\bibnamefont
  {Verstraete}}, \bibinfo {author} {\bibfnamefont {M~M}\ \bibnamefont {Wolf}},
  \ and\ \bibinfo {author} {\bibfnamefont {J~I}\ \bibnamefont {Cirac}},\
  }\bibfield  {title} {\enquote {\bibinfo {title} {{Matrix Product State
  Representations}},}\ }\href {https://dl.acm.org/doi/10.5555/2011832.2011833}
  {\bibfield  {journal} {\bibinfo  {journal} {Quantum Info. Comput.}\ }\textbf
  {\bibinfo {volume} {7}},\ \bibinfo {pages} {401--430} (\bibinfo {year}
  {2007})}\BibitemShut {NoStop}%
\bibitem [{\citenamefont {Verstraete}\ and\ \citenamefont
  {Cirac}()}]{verstraete2004renormalization}%
  \BibitemOpen
  \bibfield  {author} {\bibinfo {author} {\bibfnamefont {Frank}\ \bibnamefont
  {Verstraete}}\ and\ \bibinfo {author} {\bibfnamefont {J~Ignacio}\
  \bibnamefont {Cirac}},\ }\bibfield  {title} {\enquote {\bibinfo {title}
  {{Renormalization algorithms for quantum-many body systems in two and higher
  dimensions}},}\ }\href {https://arxiv.org/abs/cond-mat/0407066} {\ }\Eprint
  {http://arxiv.org/abs/0407066} {arXiv:0407066 [cond-mat]} \BibitemShut
  {NoStop}%
\bibitem [{\citenamefont {Eisert}\ \emph {et~al.}(2010)\citenamefont {Eisert},
  \citenamefont {Cramer},\ and\ \citenamefont {Plenio}}]{eisert2010colloquium}%
  \BibitemOpen
  \bibfield  {author} {\bibinfo {author} {\bibfnamefont {Jens}\ \bibnamefont
  {Eisert}}, \bibinfo {author} {\bibfnamefont {Marcus}\ \bibnamefont {Cramer}},
  \ and\ \bibinfo {author} {\bibfnamefont {Martin~B}\ \bibnamefont {Plenio}},\
  }\bibfield  {title} {\enquote {\bibinfo {title} {{Colloquium: Area laws for
  the entanglement entropy}},}\ }\href
  {https://journals.aps.org/rmp/abstract/10.1103/RevModPhys.82.277} {\bibfield
  {journal} {\bibinfo  {journal} {Reviews of modern physics}\ }\textbf
  {\bibinfo {volume} {82}},\ \bibinfo {pages} {277} (\bibinfo {year}
  {2010})}\BibitemShut {NoStop}%
\bibitem [{\citenamefont {Cirac}\ \emph {et~al.}(2021)\citenamefont {Cirac},
  \citenamefont {Perez-Garcia}, \citenamefont {Schuch},\ and\ \citenamefont
  {Verstraete}}]{cirac2021matrix}%
  \BibitemOpen
  \bibfield  {author} {\bibinfo {author} {\bibfnamefont {J~Ignacio}\
  \bibnamefont {Cirac}}, \bibinfo {author} {\bibfnamefont {David}\ \bibnamefont
  {Perez-Garcia}}, \bibinfo {author} {\bibfnamefont {Norbert}\ \bibnamefont
  {Schuch}}, \ and\ \bibinfo {author} {\bibfnamefont {Frank}\ \bibnamefont
  {Verstraete}},\ }\bibfield  {title} {\enquote {\bibinfo {title} {{Matrix
  product states and projected entangled pair states: Concepts, symmetries,
  theorems}},}\ }\href
  {https://journals.aps.org/rmp/abstract/10.1103/RevModPhys.93.045003}
  {\bibfield  {journal} {\bibinfo  {journal} {Reviews of Modern Physics}\
  }\textbf {\bibinfo {volume} {93}},\ \bibinfo {pages} {45003} (\bibinfo {year}
  {2021})}\BibitemShut {NoStop}%
\bibitem [{\citenamefont {Briegel}\ and\ \citenamefont
  {Raussendorf}(2001)}]{Briegel2001}%
  \BibitemOpen
  \bibfield  {author} {\bibinfo {author} {\bibfnamefont {Hans~J.}\ \bibnamefont
  {Briegel}}\ and\ \bibinfo {author} {\bibfnamefont {Robert}\ \bibnamefont
  {Raussendorf}},\ }\bibfield  {title} {\enquote {\bibinfo {title} {{Persistent
  entanglement in arrays of interacting particles}},}\ }\href {\doibase
  10.1103/PhysRevLett.86.910} {\bibfield  {journal} {\bibinfo  {journal}
  {Physical Review Letters}\ }\textbf {\bibinfo {volume} {86}},\ \bibinfo
  {pages} {910--913} (\bibinfo {year} {2001})}\BibitemShut {NoStop}%
\bibitem [{\citenamefont {Kitaev}(2003)}]{Kitaev2003}%
  \BibitemOpen
  \bibfield  {author} {\bibinfo {author} {\bibfnamefont {A~Yu}\ \bibnamefont
  {Kitaev}},\ }\bibfield  {title} {\enquote {\bibinfo {title} {{Fault-tolerant
  quantum computation by anyons}},}\ }\href
  {https://www.sciencedirect.com/science/article/abs/pii/S0003491602000180}
  {\bibfield  {journal} {\bibinfo  {journal} {Annals of Physics}\ }\textbf
  {\bibinfo {volume} {303}},\ \bibinfo {pages} {2--30} (\bibinfo {year}
  {2003})}\BibitemShut {NoStop}%
\bibitem [{\citenamefont {Greenberger}\ \emph {et~al.}(1989)\citenamefont
  {Greenberger}, \citenamefont {Horne},\ and\ \citenamefont
  {Zeilinger}}]{greenberger1989going}%
  \BibitemOpen
  \bibfield  {author} {\bibinfo {author} {\bibfnamefont {Daniel~M}\
  \bibnamefont {Greenberger}}, \bibinfo {author} {\bibfnamefont {Michael~A}\
  \bibnamefont {Horne}}, \ and\ \bibinfo {author} {\bibfnamefont {Anton}\
  \bibnamefont {Zeilinger}},\ }\bibfield  {title} {\enquote {\bibinfo {title}
  {{Going beyond Bell's theorem}},}\ }in\ \href
  {https://link.springer.com/chapter/10.1007/978-94-017-0849-4_10} {\emph
  {\bibinfo {booktitle} {Bell's theorem, quantum theory and conceptions of the
  universe}}}\ (\bibinfo  {publisher} {Springer},\ \bibinfo {year} {1989})\
  pp.\ \bibinfo {pages} {69--72}\BibitemShut {NoStop}%
\bibitem [{\citenamefont {D{\"{u}}r}\ \emph {et~al.}(2000)\citenamefont
  {D{\"{u}}r}, \citenamefont {Vidal},\ and\ \citenamefont
  {Cirac}}]{dur2000three}%
  \BibitemOpen
  \bibfield  {author} {\bibinfo {author} {\bibfnamefont {Wolfgang}\
  \bibnamefont {D{\"{u}}r}}, \bibinfo {author} {\bibfnamefont {Guifre}\
  \bibnamefont {Vidal}}, \ and\ \bibinfo {author} {\bibfnamefont {J~Ignacio}\
  \bibnamefont {Cirac}},\ }\bibfield  {title} {\enquote {\bibinfo {title}
  {{Three qubits can be entangled in two inequivalent ways}},}\ }\href
  {https://journals.aps.org/pra/abstract/10.1103/PhysRevA.62.062314} {\bibfield
   {journal} {\bibinfo  {journal} {Physical Review A}\ }\textbf {\bibinfo
  {volume} {62}},\ \bibinfo {pages} {62314} (\bibinfo {year}
  {2000})}\BibitemShut {NoStop}%
\bibitem [{\citenamefont {Affleck}\ \emph {et~al.}(1987)\citenamefont
  {Affleck}, \citenamefont {Kennedy}, \citenamefont {Lieb},\ and\ \citenamefont
  {Tasaki}}]{Affleck1987}%
  \BibitemOpen
  \bibfield  {author} {\bibinfo {author} {\bibfnamefont {Ian}\ \bibnamefont
  {Affleck}}, \bibinfo {author} {\bibfnamefont {Tom}\ \bibnamefont {Kennedy}},
  \bibinfo {author} {\bibfnamefont {Elliott~H.}\ \bibnamefont {Lieb}}, \ and\
  \bibinfo {author} {\bibfnamefont {Hal}\ \bibnamefont {Tasaki}},\ }\bibfield
  {title} {\enquote {\bibinfo {title} {{Rigorous results on valence-bond ground
  states in antiferromagnets}},}\ }\href {\doibase 10.1103/PhysRevLett.59.799}
  {\bibfield  {journal} {\bibinfo  {journal} {Physical Review Letters}\
  }\textbf {\bibinfo {volume} {59}},\ \bibinfo {pages} {799--802} (\bibinfo
  {year} {1987})}\BibitemShut {NoStop}%
\bibitem [{\citenamefont {Affleck}\ \emph {et~al.}(1988)\citenamefont
  {Affleck}, \citenamefont {Kennedy}, \citenamefont {Lieb},\ and\ \citenamefont
  {Tasaki}}]{Affleck1988}%
  \BibitemOpen
  \bibfield  {author} {\bibinfo {author} {\bibfnamefont {Ian}\ \bibnamefont
  {Affleck}}, \bibinfo {author} {\bibfnamefont {Tom}\ \bibnamefont {Kennedy}},
  \bibinfo {author} {\bibfnamefont {Elliott~H.}\ \bibnamefont {Lieb}}, \ and\
  \bibinfo {author} {\bibfnamefont {Hal}\ \bibnamefont {Tasaki}},\ }\bibfield
  {title} {\enquote {\bibinfo {title} {{Valence bond ground states in isotropic
  quantum antiferromagnets}},}\ }\href {\doibase 10.1007/BF01218021} {\bibfield
   {journal} {\bibinfo  {journal} {Communications in Mathematical Physics}\
  }\textbf {\bibinfo {volume} {115}},\ \bibinfo {pages} {477--528} (\bibinfo
  {year} {1988})}\BibitemShut {NoStop}%
\bibitem [{\citenamefont {Anderson}(1987)}]{anderson1987resonating}%
  \BibitemOpen
  \bibfield  {author} {\bibinfo {author} {\bibfnamefont {Philip~W}\
  \bibnamefont {Anderson}},\ }\bibfield  {title} {\enquote {\bibinfo {title}
  {{The resonating valence bond state in La2CuO4 and superconductivity}},}\
  }\href@noop {} {\bibfield  {journal} {\bibinfo  {journal} {science}\ }\textbf
  {\bibinfo {volume} {235}},\ \bibinfo {pages} {1196--1198} (\bibinfo {year}
  {1987})}\BibitemShut {NoStop}%
\bibitem [{\citenamefont {Levin}\ and\ \citenamefont
  {Wen}(2005)}]{levin2005string}%
  \BibitemOpen
  \bibfield  {author} {\bibinfo {author} {\bibfnamefont {Michael~A}\
  \bibnamefont {Levin}}\ and\ \bibinfo {author} {\bibfnamefont {Xiao-Gang}\
  \bibnamefont {Wen}},\ }\bibfield  {title} {\enquote {\bibinfo {title}
  {{String-net condensation: A physical mechanism for topological phases}},}\
  }\href {https://journals.aps.org/prb/abstract/10.1103/PhysRevB.71.045110}
  {\bibfield  {journal} {\bibinfo  {journal} {Physical Review B}\ }\textbf
  {\bibinfo {volume} {71}},\ \bibinfo {pages} {45110} (\bibinfo {year}
  {2005})}\BibitemShut {NoStop}%
\bibitem [{\citenamefont {Briegel}\ \emph {et~al.}(2009)\citenamefont
  {Briegel}, \citenamefont {Browne}, \citenamefont {D{\"{u}}r}, \citenamefont
  {Raussendorf},\ and\ \citenamefont {den Nest}}]{briegel2009measurement}%
  \BibitemOpen
  \bibfield  {author} {\bibinfo {author} {\bibfnamefont {Hans~J}\ \bibnamefont
  {Briegel}}, \bibinfo {author} {\bibfnamefont {David~E}\ \bibnamefont
  {Browne}}, \bibinfo {author} {\bibfnamefont {Wolfgang}\ \bibnamefont
  {D{\"{u}}r}}, \bibinfo {author} {\bibfnamefont {Robert}\ \bibnamefont
  {Raussendorf}}, \ and\ \bibinfo {author} {\bibfnamefont {Maarten}\
  \bibnamefont {den Nest}},\ }\bibfield  {title} {\enquote {\bibinfo {title}
  {{Measurement-based quantum computation}},}\ }\href
  {https://www.nature.com/articles/nphys1157} {\bibfield  {journal} {\bibinfo
  {journal} {Nature Physics}\ }\textbf {\bibinfo {volume} {5}},\ \bibinfo
  {pages} {19--26} (\bibinfo {year} {2009})}\BibitemShut {NoStop}%
\bibitem [{\citenamefont {Jarzyna}\ and\ \citenamefont
  {Demkowicz-Dobrza{\'{n}}ski}(2013)}]{Jarzyna2013}%
  \BibitemOpen
  \bibfield  {author} {\bibinfo {author} {\bibfnamefont {Marcin}\ \bibnamefont
  {Jarzyna}}\ and\ \bibinfo {author} {\bibfnamefont {Rafa{\l}}\ \bibnamefont
  {Demkowicz-Dobrza{\'{n}}ski}},\ }\bibfield  {title} {\enquote {\bibinfo
  {title} {{Matrix product states for quantum metrology}},}\ }\href {\doibase
  10.1103/PhysRevLett.110.240405} {\bibfield  {journal} {\bibinfo  {journal}
  {Physical Review Letters}\ }\textbf {\bibinfo {volume} {110}},\ \bibinfo
  {pages} {1--5} (\bibinfo {year} {2013})}\BibitemShut {NoStop}%
\bibitem [{\citenamefont {Azuma}\ \emph {et~al.}(2015)\citenamefont {Azuma},
  \citenamefont {Tamaki},\ and\ \citenamefont {Lo}}]{Azuma2015}%
  \BibitemOpen
  \bibfield  {author} {\bibinfo {author} {\bibfnamefont {Koji}\ \bibnamefont
  {Azuma}}, \bibinfo {author} {\bibfnamefont {Kiyoshi}\ \bibnamefont {Tamaki}},
  \ and\ \bibinfo {author} {\bibfnamefont {Hoi~Kwong}\ \bibnamefont {Lo}},\
  }\bibfield  {title} {\enquote {\bibinfo {title} {{All-photonic quantum
  repeaters}},}\ }\href {\doibase 10.1038/ncomms7787} {\bibfield  {journal}
  {\bibinfo  {journal} {Nature Communications}\ }\textbf {\bibinfo {volume}
  {6}} (\bibinfo {year} {2015}),\ 10.1038/ncomms7787}\BibitemShut {NoStop}%
\bibitem [{\citenamefont {Huggins}\ \emph {et~al.}(2019)\citenamefont
  {Huggins}, \citenamefont {Patil}, \citenamefont {Mitchell}, \citenamefont
  {{Birgitta Whaley}},\ and\ \citenamefont {{Miles
  Stoudenmire}}}]{Huggins2019}%
  \BibitemOpen
  \bibfield  {author} {\bibinfo {author} {\bibfnamefont {William}\ \bibnamefont
  {Huggins}}, \bibinfo {author} {\bibfnamefont {Piyush}\ \bibnamefont {Patil}},
  \bibinfo {author} {\bibfnamefont {Bradley}\ \bibnamefont {Mitchell}},
  \bibinfo {author} {\bibfnamefont {K.}~\bibnamefont {{Birgitta Whaley}}}, \
  and\ \bibinfo {author} {\bibfnamefont {E.}~\bibnamefont {{Miles
  Stoudenmire}}},\ }\bibfield  {title} {\enquote {\bibinfo {title} {{Towards
  quantum machine learning with tensor networks}},}\ }\href {\doibase
  10.1088/2058-9565/aaea94} {\bibfield  {journal} {\bibinfo  {journal} {Quantum
  Science and Technology}\ }\textbf {\bibinfo {volume} {4}} (\bibinfo {year}
  {2019}),\ 10.1088/2058-9565/aaea94}\BibitemShut {NoStop}%
\bibitem [{\citenamefont {Sch{\"{o}}n}\ \emph {et~al.}(2005)\citenamefont
  {Sch{\"{o}}n}, \citenamefont {Solano}, \citenamefont {Verstraete},
  \citenamefont {Cirac},\ and\ \citenamefont {Wolf}}]{Schon2005}%
  \BibitemOpen
  \bibfield  {author} {\bibinfo {author} {\bibfnamefont {C.}~\bibnamefont
  {Sch{\"{o}}n}}, \bibinfo {author} {\bibfnamefont {E.}~\bibnamefont {Solano}},
  \bibinfo {author} {\bibfnamefont {F.}~\bibnamefont {Verstraete}}, \bibinfo
  {author} {\bibfnamefont {J.~I.}\ \bibnamefont {Cirac}}, \ and\ \bibinfo
  {author} {\bibfnamefont {M.~M.}\ \bibnamefont {Wolf}},\ }\bibfield  {title}
  {\enquote {\bibinfo {title} {{Sequential generation of entangled multiqubit
  states}},}\ }\href {\doibase 10.1103/PhysRevLett.95.110503} {\bibfield
  {journal} {\bibinfo  {journal} {Physical Review Letters}\ }\textbf {\bibinfo
  {volume} {95}},\ \bibinfo {pages} {1--4} (\bibinfo {year}
  {2005})}\BibitemShut {NoStop}%
\bibitem [{\citenamefont {Schuch}\ \emph {et~al.}(2007)\citenamefont {Schuch},
  \citenamefont {Wolf}, \citenamefont {Verstraete},\ and\ \citenamefont
  {Cirac}}]{Schuch2007}%
  \BibitemOpen
  \bibfield  {author} {\bibinfo {author} {\bibfnamefont {Norbert}\ \bibnamefont
  {Schuch}}, \bibinfo {author} {\bibfnamefont {Michael~M.}\ \bibnamefont
  {Wolf}}, \bibinfo {author} {\bibfnamefont {Frank}\ \bibnamefont
  {Verstraete}}, \ and\ \bibinfo {author} {\bibfnamefont {J.~Ignacio}\
  \bibnamefont {Cirac}},\ }\bibfield  {title} {\enquote {\bibinfo {title}
  {{Computational complexity of projected entangled pair states}},}\ }\href
  {\doibase 10.1103/PhysRevLett.98.140506} {\bibfield  {journal} {\bibinfo
  {journal} {Physical Review Letters}\ }\textbf {\bibinfo {volume} {98}},\
  \bibinfo {pages} {1--4} (\bibinfo {year} {2007})}\BibitemShut {NoStop}%
\bibitem [{\citenamefont {Ba{\~{n}}uls}\ \emph {et~al.}(2008)\citenamefont
  {Ba{\~{n}}uls}, \citenamefont {P{\'{e}}rez-Garc{\'{i}}a}, \citenamefont
  {Wolf}, \citenamefont {Verstraete},\ and\ \citenamefont
  {Cirac}}]{Banuls2008}%
  \BibitemOpen
  \bibfield  {author} {\bibinfo {author} {\bibfnamefont {M.~C.}\ \bibnamefont
  {Ba{\~{n}}uls}}, \bibinfo {author} {\bibfnamefont {D.}~\bibnamefont
  {P{\'{e}}rez-Garc{\'{i}}a}}, \bibinfo {author} {\bibfnamefont {M.~M.}\
  \bibnamefont {Wolf}}, \bibinfo {author} {\bibfnamefont {F.}~\bibnamefont
  {Verstraete}}, \ and\ \bibinfo {author} {\bibfnamefont {J.~I.}\ \bibnamefont
  {Cirac}},\ }\bibfield  {title} {\enquote {\bibinfo {title} {{Sequentially
  generated states for the study of two-dimensional systems}},}\ }\href
  {\doibase 10.1103/PhysRevA.77.052306} {\bibfield  {journal} {\bibinfo
  {journal} {Physical Review A}\ }\textbf {\bibinfo {volume} {77}},\ \bibinfo
  {pages} {1--9} (\bibinfo {year} {2008})}\BibitemShut {NoStop}%
\bibitem [{\citenamefont {Pichler}\ \emph {et~al.}(2017)\citenamefont
  {Pichler}, \citenamefont {Choi}, \citenamefont {Zoller},\ and\ \citenamefont
  {Lukin}}]{Pichler2017}%
  \BibitemOpen
  \bibfield  {author} {\bibinfo {author} {\bibfnamefont {Hannes}\ \bibnamefont
  {Pichler}}, \bibinfo {author} {\bibfnamefont {Soonwon}\ \bibnamefont {Choi}},
  \bibinfo {author} {\bibfnamefont {Peter}\ \bibnamefont {Zoller}}, \ and\
  \bibinfo {author} {\bibfnamefont {Mikhail~D.}\ \bibnamefont {Lukin}},\
  }\bibfield  {title} {\enquote {\bibinfo {title} {{Universal photonic quantum
  computation via time-delayed feedback}},}\ }\href {\doibase
  10.1073/pnas.1711003114} {\bibfield  {journal} {\bibinfo  {journal}
  {Proceedings of the National Academy of Sciences}\ }\textbf {\bibinfo
  {volume} {114}},\ \bibinfo {pages} {11362--11367} (\bibinfo {year}
  {2017})}\BibitemShut {NoStop}%
\bibitem [{\citenamefont {Zaletel}\ and\ \citenamefont
  {Pollmann}(2020)}]{Zaletel2020}%
  \BibitemOpen
  \bibfield  {author} {\bibinfo {author} {\bibfnamefont {Michael~P.}\
  \bibnamefont {Zaletel}}\ and\ \bibinfo {author} {\bibfnamefont {Frank}\
  \bibnamefont {Pollmann}},\ }\bibfield  {title} {\enquote {\bibinfo {title}
  {{Isometric Tensor Network States in Two Dimensions}},}\ }\href {\doibase
  10.1103/PhysRevLett.124.037201} {\bibfield  {journal} {\bibinfo  {journal}
  {Physical Review Letters}\ }\textbf {\bibinfo {volume} {124}},\ \bibinfo
  {pages} {37201} (\bibinfo {year} {2020})}\BibitemShut {NoStop}%
\bibitem [{\citenamefont {Wei}\ \emph {et~al.}(2022)\citenamefont {Wei},
  \citenamefont {Malz},\ and\ \citenamefont {Cirac}}]{zypp}%
  \BibitemOpen
  \bibfield  {author} {\bibinfo {author} {\bibfnamefont {Zhi~Yuan}\
  \bibnamefont {Wei}}, \bibinfo {author} {\bibfnamefont {Daniel}\ \bibnamefont
  {Malz}}, \ and\ \bibinfo {author} {\bibfnamefont {J~Ignacio}\ \bibnamefont
  {Cirac}},\ }\bibfield  {title} {\enquote {\bibinfo {title} {{Sequential
  Generation of Projected Entangled-Pair States}},}\ }\href {\doibase
  10.1103/PhysRevLett.128.010607} {\bibfield  {journal} {\bibinfo  {journal}
  {Physical Review Letters}\ }\textbf {\bibinfo {volume} {128}},\ \bibinfo
  {pages} {1--14} (\bibinfo {year} {2022})}\BibitemShut {NoStop}%
\bibitem [{\citenamefont {Schwartz}\ \emph {et~al.}(2016)\citenamefont
  {Schwartz}, \citenamefont {Cogan}, \citenamefont {Schmidgall}, \citenamefont
  {Don}, \citenamefont {Gantz}, \citenamefont {Kenneth}, \citenamefont
  {Lindner},\ and\ \citenamefont {Gershoni}}]{schwartz2016}%
  \BibitemOpen
  \bibfield  {author} {\bibinfo {author} {\bibfnamefont {I.}~\bibnamefont
  {Schwartz}}, \bibinfo {author} {\bibfnamefont {D.}~\bibnamefont {Cogan}},
  \bibinfo {author} {\bibfnamefont {E.~R.}\ \bibnamefont {Schmidgall}},
  \bibinfo {author} {\bibfnamefont {Y.}~\bibnamefont {Don}}, \bibinfo {author}
  {\bibfnamefont {L.}~\bibnamefont {Gantz}}, \bibinfo {author} {\bibfnamefont
  {O.}~\bibnamefont {Kenneth}}, \bibinfo {author} {\bibfnamefont {N.~H.}\
  \bibnamefont {Lindner}}, \ and\ \bibinfo {author} {\bibfnamefont
  {D.}~\bibnamefont {Gershoni}},\ }\bibfield  {title} {\enquote {\bibinfo
  {title} {{Deterministic generation of a cluster state of entangled
  photons}},}\ }\href {\doibase 10.1126/science.aah4758} {\bibfield  {journal}
  {\bibinfo  {journal} {Science}\ }\textbf {\bibinfo {volume} {354}},\ \bibinfo
  {pages} {434--437} (\bibinfo {year} {2016})}\BibitemShut {NoStop}%
\bibitem [{\citenamefont {Besse}\ \emph {et~al.}(2020)\citenamefont {Besse},
  \citenamefont {Reuer}, \citenamefont {Collodo}, \citenamefont {Wulff},
  \citenamefont {Wernli}, \citenamefont {Copetudo}, \citenamefont {Malz},
  \citenamefont {Magnard}, \citenamefont {Akin}, \citenamefont {Gabureac},
  \citenamefont {Norris}, \citenamefont {Cirac}, \citenamefont {Wallraff},\
  and\ \citenamefont {Eichler}}]{Besse2020}%
  \BibitemOpen
  \bibfield  {author} {\bibinfo {author} {\bibfnamefont {Jean-Claude}\
  \bibnamefont {Besse}}, \bibinfo {author} {\bibfnamefont {Kevin}\ \bibnamefont
  {Reuer}}, \bibinfo {author} {\bibfnamefont {Michele~C.}\ \bibnamefont
  {Collodo}}, \bibinfo {author} {\bibfnamefont {Arne}\ \bibnamefont {Wulff}},
  \bibinfo {author} {\bibfnamefont {Lucien}\ \bibnamefont {Wernli}}, \bibinfo
  {author} {\bibfnamefont {Adrian}\ \bibnamefont {Copetudo}}, \bibinfo {author}
  {\bibfnamefont {Daniel}\ \bibnamefont {Malz}}, \bibinfo {author}
  {\bibfnamefont {Paul}\ \bibnamefont {Magnard}}, \bibinfo {author}
  {\bibfnamefont {Abdulkadir}\ \bibnamefont {Akin}}, \bibinfo {author}
  {\bibfnamefont {Mihai}\ \bibnamefont {Gabureac}}, \bibinfo {author}
  {\bibfnamefont {Graham~J.}\ \bibnamefont {Norris}}, \bibinfo {author}
  {\bibfnamefont {J.~Ignacio}\ \bibnamefont {Cirac}}, \bibinfo {author}
  {\bibfnamefont {Andreas}\ \bibnamefont {Wallraff}}, \ and\ \bibinfo {author}
  {\bibfnamefont {Christopher}\ \bibnamefont {Eichler}},\ }\bibfield  {title}
  {\enquote {\bibinfo {title} {{Realizing a deterministic source of
  multipartite-entangled photonic qubits}},}\ }\href {\doibase
  10.1038/s41467-020-18635-x} {\bibfield  {journal} {\bibinfo  {journal}
  {Nature Communications}\ }\textbf {\bibinfo {volume} {11}},\ \bibinfo {pages}
  {1--6} (\bibinfo {year} {2020})}\BibitemShut {NoStop}%
\bibitem [{\citenamefont {J.}\ \emph {et~al.}(2021)\citenamefont {J.},
  \citenamefont {Y.-J}, \citenamefont {A.}, \citenamefont {C.}, \citenamefont
  {M.}, \citenamefont {C.}, \citenamefont {Z.}, \citenamefont {C.},
  \citenamefont {X.}, \citenamefont {A.}, \citenamefont {C.}, \citenamefont
  {I.}, \citenamefont {F.}, \citenamefont {K.}, \citenamefont {J.},
  \citenamefont {R.}, \citenamefont {C.}, \citenamefont {R.}, \citenamefont
  {J.}, \citenamefont {A.}, \citenamefont {A.}, \citenamefont {M.},
  \citenamefont {B.}, \citenamefont {A.}, \citenamefont {B.}, \citenamefont
  {N.}, \citenamefont {B.}, \citenamefont {R.}, \citenamefont {W.},
  \citenamefont {S.}, \citenamefont {R.}, \citenamefont {D.}, \citenamefont
  {C.}, \citenamefont {L.}, \citenamefont {E.}, \citenamefont {G.},
  \citenamefont {B.}, \citenamefont {M.}, \citenamefont {A.}, \citenamefont
  {A.}, \citenamefont {P.}, \citenamefont {D.}, \citenamefont {J.},
  \citenamefont {S.}, \citenamefont {T.}, \citenamefont {J.}, \citenamefont
  {B.}, \citenamefont {V.}, \citenamefont {E.}, \citenamefont {Z.},
  \citenamefont {D.}, \citenamefont {K.}, \citenamefont {T.}, \citenamefont
  {S.}, \citenamefont {V.}, \citenamefont {N.}, \citenamefont {F.},
  \citenamefont {D.}, \citenamefont {P.}, \citenamefont {A.}, \citenamefont
  {E.}, \citenamefont {O.}, \citenamefont {R.}, \citenamefont {M.},
  \citenamefont {C.}, \citenamefont {M.}, \citenamefont {S.}, \citenamefont
  {W.}, \citenamefont {J.}, \citenamefont {O.}, \citenamefont {M.},
  \citenamefont {C.}, \citenamefont {Y.}, \citenamefont {E.}, \citenamefont
  {A.}, \citenamefont {B.}, \citenamefont {A.}, \citenamefont {C.},
  \citenamefont {D.}, \citenamefont {V.}, \citenamefont {D.}, \citenamefont
  {M.}, \citenamefont {B.}, \citenamefont {C.}, \citenamefont {Z.},
  \citenamefont {P.}, \citenamefont {J.}, \citenamefont {A.}, \citenamefont
  {H.}, \citenamefont {S.}, \citenamefont {A.}, \citenamefont {Y.},
  \citenamefont {J.}, \citenamefont {V.}, \citenamefont {A.}, \citenamefont
  {M.}, \citenamefont {F.},\ and\ \citenamefont {P.}}]{satzinger2021}%
  \BibitemOpen
  \bibfield  {author} {\bibinfo {author} {\bibfnamefont {Satzinger~K}\
  \bibnamefont {J.}}, \bibinfo {author} {\bibfnamefont {Liu}\ \bibnamefont
  {Y.-J}}, \bibinfo {author} {\bibfnamefont {Smith}\ \bibnamefont {A.}},
  \bibinfo {author} {\bibfnamefont {Knapp}\ \bibnamefont {C.}}, \bibinfo
  {author} {\bibfnamefont {Newman}\ \bibnamefont {M.}}, \bibinfo {author}
  {\bibfnamefont {Jones}\ \bibnamefont {C.}}, \bibinfo {author} {\bibfnamefont
  {Chen}\ \bibnamefont {Z.}}, \bibinfo {author} {\bibfnamefont {Quintana}\
  \bibnamefont {C.}}, \bibinfo {author} {\bibfnamefont {Mi}~\bibnamefont {X.}},
  \bibinfo {author} {\bibfnamefont {Dunsworth}\ \bibnamefont {A.}}, \bibinfo
  {author} {\bibfnamefont {Gidney}\ \bibnamefont {C.}}, \bibinfo {author}
  {\bibfnamefont {Aleiner}\ \bibnamefont {I.}}, \bibinfo {author}
  {\bibfnamefont {Arute}\ \bibnamefont {F.}}, \bibinfo {author} {\bibfnamefont
  {Arya}\ \bibnamefont {K.}}, \bibinfo {author} {\bibfnamefont {Atalaya}\
  \bibnamefont {J.}}, \bibinfo {author} {\bibfnamefont {Babbush}\ \bibnamefont
  {R.}}, \bibinfo {author} {\bibfnamefont {Bardin~J}\ \bibnamefont {C.}},
  \bibinfo {author} {\bibfnamefont {Barends}\ \bibnamefont {R.}}, \bibinfo
  {author} {\bibfnamefont {Basso}\ \bibnamefont {J.}}, \bibinfo {author}
  {\bibfnamefont {Bengtsson}\ \bibnamefont {A.}}, \bibinfo {author}
  {\bibfnamefont {Bilmes}\ \bibnamefont {A.}}, \bibinfo {author} {\bibfnamefont
  {Broughton}\ \bibnamefont {M.}}, \bibinfo {author} {\bibfnamefont
  {Buckley~B}\ \bibnamefont {B.}}, \bibinfo {author} {\bibfnamefont {Buell~D}\
  \bibnamefont {A.}}, \bibinfo {author} {\bibfnamefont {Burkett}\ \bibnamefont
  {B.}}, \bibinfo {author} {\bibfnamefont {Bushnell}\ \bibnamefont {N.}},
  \bibinfo {author} {\bibfnamefont {Chiaro}\ \bibnamefont {B.}}, \bibinfo
  {author} {\bibfnamefont {Collins}\ \bibnamefont {R.}}, \bibinfo {author}
  {\bibfnamefont {Courtney}\ \bibnamefont {W.}}, \bibinfo {author}
  {\bibfnamefont {Demura}\ \bibnamefont {S.}}, \bibinfo {author} {\bibfnamefont
  {Derk~A}\ \bibnamefont {R.}}, \bibinfo {author} {\bibfnamefont {Eppens}\
  \bibnamefont {D.}}, \bibinfo {author} {\bibfnamefont {Erickson}\ \bibnamefont
  {C.}}, \bibinfo {author} {\bibfnamefont {Faoro}\ \bibnamefont {L.}}, \bibinfo
  {author} {\bibfnamefont {Farhi}\ \bibnamefont {E.}}, \bibinfo {author}
  {\bibfnamefont {Fowler~A}\ \bibnamefont {G.}}, \bibinfo {author}
  {\bibfnamefont {Foxen}\ \bibnamefont {B.}}, \bibinfo {author} {\bibfnamefont
  {Giustina}\ \bibnamefont {M.}}, \bibinfo {author} {\bibfnamefont {Greene}\
  \bibnamefont {A.}}, \bibinfo {author} {\bibfnamefont {Gross~J}\ \bibnamefont
  {A.}}, \bibinfo {author} {\bibfnamefont {Harrigan~M}\ \bibnamefont {P.}},
  \bibinfo {author} {\bibfnamefont {Harrington~S}\ \bibnamefont {D.}}, \bibinfo
  {author} {\bibfnamefont {Hilton}\ \bibnamefont {J.}}, \bibinfo {author}
  {\bibfnamefont {Hong}\ \bibnamefont {S.}}, \bibinfo {author} {\bibfnamefont
  {Huang}\ \bibnamefont {T.}}, \bibinfo {author} {\bibfnamefont {Huggins~W}\
  \bibnamefont {J.}}, \bibinfo {author} {\bibfnamefont {Ioffe~L}\ \bibnamefont
  {B.}}, \bibinfo {author} {\bibfnamefont {Isakov~S}\ \bibnamefont {V.}},
  \bibinfo {author} {\bibfnamefont {Jeffrey}\ \bibnamefont {E.}}, \bibinfo
  {author} {\bibfnamefont {Jiang}\ \bibnamefont {Z.}}, \bibinfo {author}
  {\bibfnamefont {Kafri}\ \bibnamefont {D.}}, \bibinfo {author} {\bibfnamefont
  {Kechedzhi}\ \bibnamefont {K.}}, \bibinfo {author} {\bibfnamefont {Khattar}\
  \bibnamefont {T.}}, \bibinfo {author} {\bibfnamefont {Kim}\ \bibnamefont
  {S.}}, \bibinfo {author} {\bibfnamefont {Klimov~P}\ \bibnamefont {V.}},
  \bibinfo {author} {\bibfnamefont {Korotkov~A}\ \bibnamefont {N.}}, \bibinfo
  {author} {\bibfnamefont {Kostritsa}\ \bibnamefont {F.}}, \bibinfo {author}
  {\bibfnamefont {Landhuis}\ \bibnamefont {D.}}, \bibinfo {author}
  {\bibfnamefont {Laptev}\ \bibnamefont {P.}}, \bibinfo {author} {\bibfnamefont
  {Locharla}\ \bibnamefont {A.}}, \bibinfo {author} {\bibfnamefont {Lucero}\
  \bibnamefont {E.}}, \bibinfo {author} {\bibfnamefont {Martin}\ \bibnamefont
  {O.}}, \bibinfo {author} {\bibfnamefont {McClean~J}\ \bibnamefont {R.}},
  \bibinfo {author} {\bibfnamefont {McEwen}\ \bibnamefont {M.}}, \bibinfo
  {author} {\bibfnamefont {Miao~K}\ \bibnamefont {C.}}, \bibinfo {author}
  {\bibfnamefont {Mohseni}\ \bibnamefont {M.}}, \bibinfo {author}
  {\bibfnamefont {Montazeri}\ \bibnamefont {S.}}, \bibinfo {author}
  {\bibfnamefont {Mruczkiewicz}\ \bibnamefont {W.}}, \bibinfo {author}
  {\bibfnamefont {Mutus}\ \bibnamefont {J.}}, \bibinfo {author} {\bibfnamefont
  {Naaman}\ \bibnamefont {O.}}, \bibinfo {author} {\bibfnamefont {Neeley}\
  \bibnamefont {M.}}, \bibinfo {author} {\bibfnamefont {Neill}\ \bibnamefont
  {C.}}, \bibinfo {author} {\bibfnamefont {Niu~M}\ \bibnamefont {Y.}}, \bibinfo
  {author} {\bibfnamefont {O'Brien~T}\ \bibnamefont {E.}}, \bibinfo {author}
  {\bibfnamefont {Opremcak}\ \bibnamefont {A.}}, \bibinfo {author}
  {\bibfnamefont {Pat{\'{o}}}\ \bibnamefont {B.}}, \bibinfo {author}
  {\bibfnamefont {Petukhov}\ \bibnamefont {A.}}, \bibinfo {author}
  {\bibfnamefont {Rubin~N}\ \bibnamefont {C.}}, \bibinfo {author}
  {\bibfnamefont {Sank}\ \bibnamefont {D.}}, \bibinfo {author} {\bibfnamefont
  {Shvarts}\ \bibnamefont {V.}}, \bibinfo {author} {\bibfnamefont {Strain}\
  \bibnamefont {D.}}, \bibinfo {author} {\bibfnamefont {Szalay}\ \bibnamefont
  {M.}}, \bibinfo {author} {\bibfnamefont {Villalonga}\ \bibnamefont {B.}},
  \bibinfo {author} {\bibfnamefont {White~T}\ \bibnamefont {C.}}, \bibinfo
  {author} {\bibfnamefont {Yao}\ \bibnamefont {Z.}}, \bibinfo {author}
  {\bibfnamefont {Yeh}\ \bibnamefont {P.}}, \bibinfo {author} {\bibfnamefont
  {Yoo}\ \bibnamefont {J.}}, \bibinfo {author} {\bibfnamefont {Zalcman}\
  \bibnamefont {A.}}, \bibinfo {author} {\bibfnamefont {Neven}\ \bibnamefont
  {H.}}, \bibinfo {author} {\bibfnamefont {Boixo}\ \bibnamefont {S.}}, \bibinfo
  {author} {\bibfnamefont {Megrant}\ \bibnamefont {A.}}, \bibinfo {author}
  {\bibfnamefont {Chen}\ \bibnamefont {Y.}}, \bibinfo {author} {\bibfnamefont
  {Kelly}\ \bibnamefont {J.}}, \bibinfo {author} {\bibfnamefont {Smelyanskiy}\
  \bibnamefont {V.}}, \bibinfo {author} {\bibfnamefont {Kitaev}\ \bibnamefont
  {A.}}, \bibinfo {author} {\bibfnamefont {Knap}\ \bibnamefont {M.}}, \bibinfo
  {author} {\bibfnamefont {Pollmann}\ \bibnamefont {F.}}, \ and\ \bibinfo
  {author} {\bibfnamefont {Roushan}\ \bibnamefont {P.}},\ }\bibfield  {title}
  {\enquote {\bibinfo {title} {{Realizing topologically ordered states on a
  quantum processor}},}\ }\href {\doibase 10.1126/science.abi8378} {\bibfield
  {journal} {\bibinfo  {journal} {Science}\ }\textbf {\bibinfo {volume}
  {374}},\ \bibinfo {pages} {1237--1241} (\bibinfo {year} {2021})}\BibitemShut
  {NoStop}%
\bibitem [{\citenamefont {Smith}\ \emph {et~al.}(2022)\citenamefont {Smith},
  \citenamefont {Jobst}, \citenamefont {Green},\ and\ \citenamefont
  {Pollmann}}]{smith2022crossing}%
  \BibitemOpen
  \bibfield  {author} {\bibinfo {author} {\bibfnamefont {Adam}\ \bibnamefont
  {Smith}}, \bibinfo {author} {\bibfnamefont {Bernhard}\ \bibnamefont {Jobst}},
  \bibinfo {author} {\bibfnamefont {Andrew~G}\ \bibnamefont {Green}}, \ and\
  \bibinfo {author} {\bibfnamefont {Frank}\ \bibnamefont {Pollmann}},\
  }\bibfield  {title} {\enquote {\bibinfo {title} {{Crossing a topological
  phase transition with a quantum computer}},}\ }\href
  {https://journals.aps.org/prresearch/abstract/10.1103/PhysRevResearch.4.L022020}
  {\bibfield  {journal} {\bibinfo  {journal} {Physical Review Research}\
  }\textbf {\bibinfo {volume} {4}},\ \bibinfo {pages} {L022020} (\bibinfo
  {year} {2022})}\BibitemShut {NoStop}%
\bibitem [{\citenamefont {Albash}\ and\ \citenamefont
  {Lidar}(2018)}]{Albash2018}%
  \BibitemOpen
  \bibfield  {author} {\bibinfo {author} {\bibfnamefont {Tameem}\ \bibnamefont
  {Albash}}\ and\ \bibinfo {author} {\bibfnamefont {Daniel~A.}\ \bibnamefont
  {Lidar}},\ }\bibfield  {title} {\enquote {\bibinfo {title} {{Adiabatic
  quantum computation}},}\ }\href {\doibase 10.1103/RevModPhys.90.015002}
  {\bibfield  {journal} {\bibinfo  {journal} {Reviews of Modern Physics}\
  }\textbf {\bibinfo {volume} {90}},\ \bibinfo {pages} {15002} (\bibinfo {year}
  {2018})}\BibitemShut {NoStop}%
\bibitem [{\citenamefont {Schwarz}\ \emph {et~al.}(2012)\citenamefont
  {Schwarz}, \citenamefont {Temme},\ and\ \citenamefont
  {Verstraete}}]{Schwarz2012a}%
  \BibitemOpen
  \bibfield  {author} {\bibinfo {author} {\bibfnamefont {Martin}\ \bibnamefont
  {Schwarz}}, \bibinfo {author} {\bibfnamefont {Kristan}\ \bibnamefont
  {Temme}}, \ and\ \bibinfo {author} {\bibfnamefont {Frank}\ \bibnamefont
  {Verstraete}},\ }\bibfield  {title} {\enquote {\bibinfo {title} {{Preparing
  projected entangled pair states on a quantum computer}},}\ }\href {\doibase
  10.1103/PhysRevLett.108.110502} {\bibfield  {journal} {\bibinfo  {journal}
  {Physical Review Letters}\ }\textbf {\bibinfo {volume} {108}},\ \bibinfo
  {pages} {1--5} (\bibinfo {year} {2012})}\BibitemShut {NoStop}%
\bibitem [{\citenamefont {Schwarz}\ \emph {et~al.}(2013)\citenamefont
  {Schwarz}, \citenamefont {Temme}, \citenamefont {Verstraete}, \citenamefont
  {Perez-Garcia},\ and\ \citenamefont {Cubitt}}]{Schwarz2013}%
  \BibitemOpen
  \bibfield  {author} {\bibinfo {author} {\bibfnamefont {Martin}\ \bibnamefont
  {Schwarz}}, \bibinfo {author} {\bibfnamefont {Kristan}\ \bibnamefont
  {Temme}}, \bibinfo {author} {\bibfnamefont {Frank}\ \bibnamefont
  {Verstraete}}, \bibinfo {author} {\bibfnamefont {David}\ \bibnamefont
  {Perez-Garcia}}, \ and\ \bibinfo {author} {\bibfnamefont {Toby~S.}\
  \bibnamefont {Cubitt}},\ }\bibfield  {title} {\enquote {\bibinfo {title}
  {{Preparing topological projected entangled pair states on a quantum
  computer}},}\ }\href {\doibase 10.1103/PhysRevA.88.032321} {\bibfield
  {journal} {\bibinfo  {journal} {Physical Review A}\ }\textbf {\bibinfo
  {volume} {88}},\ \bibinfo {pages} {1--6} (\bibinfo {year}
  {2013})}\BibitemShut {NoStop}%
\bibitem [{\citenamefont {Ge}\ \emph {et~al.}(2016)\citenamefont {Ge},
  \citenamefont {Moln{\'{a}}r},\ and\ \citenamefont {Cirac}}]{Ge2016}%
  \BibitemOpen
  \bibfield  {author} {\bibinfo {author} {\bibfnamefont {Yimin}\ \bibnamefont
  {Ge}}, \bibinfo {author} {\bibfnamefont {Andr{\'{a}}s}\ \bibnamefont
  {Moln{\'{a}}r}}, \ and\ \bibinfo {author} {\bibfnamefont {J.~Ignacio}\
  \bibnamefont {Cirac}},\ }\bibfield  {title} {\enquote {\bibinfo {title}
  {{Rapid Adiabatic Preparation of Injective Projected Entangled Pair States
  and Gibbs States}},}\ }\href {\doibase 10.1103/PhysRevLett.116.080503}
  {\bibfield  {journal} {\bibinfo  {journal} {Physical Review Letters}\
  }\textbf {\bibinfo {volume} {116}},\ \bibinfo {pages} {1--5} (\bibinfo {year}
  {2016})}\BibitemShut {NoStop}%
\bibitem [{\citenamefont {Cruz}\ \emph {et~al.}(2022)\citenamefont {Cruz},
  \citenamefont {Baccari}, \citenamefont {Tura}, \citenamefont {Schuch},\ and\
  \citenamefont {Cirac}}]{cruz2022preparation}%
  \BibitemOpen
  \bibfield  {author} {\bibinfo {author} {\bibfnamefont {Esther}\ \bibnamefont
  {Cruz}}, \bibinfo {author} {\bibfnamefont {Flavio}\ \bibnamefont {Baccari}},
  \bibinfo {author} {\bibfnamefont {Jordi}\ \bibnamefont {Tura}}, \bibinfo
  {author} {\bibfnamefont {Norbert}\ \bibnamefont {Schuch}}, \ and\ \bibinfo
  {author} {\bibfnamefont {J~Ignacio}\ \bibnamefont {Cirac}},\ }\bibfield
  {title} {\enquote {\bibinfo {title} {{Preparation and verification of tensor
  network states}},}\ }\href
  {https://journals.aps.org/prresearch/abstract/10.1103/PhysRevResearch.4.023161}
  {\bibfield  {journal} {\bibinfo  {journal} {Physical Review Research}\
  }\textbf {\bibinfo {volume} {4}},\ \bibinfo {pages} {23161} (\bibinfo {year}
  {2022})}\BibitemShut {NoStop}%
\bibitem [{sup()}]{supp}%
  \BibitemOpen
  \bibfield  {title} {\enquote {\bibinfo {title} {{See Supplemental Material at
  [URL will be inserted by publisher] for more details}},}\ }\href@noop {}
  {\bibinfo  {journal} {See Supplemental Material at [URL will be inserted by
  publisher] for more details}\ }\BibitemShut {NoStop}%
\bibitem [{\citenamefont {Wolf}\ \emph {et~al.}(2006)\citenamefont {Wolf},
  \citenamefont {Ortiz}, \citenamefont {Verstraete},\ and\ \citenamefont
  {Cirac}}]{Wolf2006}%
  \BibitemOpen
\bibfield  {journal} {  }\bibfield  {author} {\bibinfo {author} {\bibfnamefont
  {Michael~M.}\ \bibnamefont {Wolf}}, \bibinfo {author} {\bibfnamefont
  {Gerardo}\ \bibnamefont {Ortiz}}, \bibinfo {author} {\bibfnamefont {Frank}\
  \bibnamefont {Verstraete}}, \ and\ \bibinfo {author} {\bibfnamefont
  {J.~Ignacio}\ \bibnamefont {Cirac}},\ }\bibfield  {title} {\enquote {\bibinfo
  {title} {{Quantum phase transitions in matrix product systems}},}\ }\href
  {\doibase 10.1103/PhysRevLett.97.110403} {\bibfield  {journal} {\bibinfo
  {journal} {Physical Review Letters}\ }\textbf {\bibinfo {volume} {97}},\
  \bibinfo {pages} {1--4} (\bibinfo {year} {2006})}\BibitemShut {NoStop}%
\bibitem [{Note1()}]{Note1}%
  \BibitemOpen
  \bibinfo {note} {Note that an asymptotic scaling of this kind is not possible
  with the adiabatic ramps we use in this paper, as shown in~\cite
  {nenciu1993linear,hagedorn2002elementary,Rezakhani2010} and the
  appendix~\cite {supp}.}\BibitemShut {Stop}%
\bibitem [{\citenamefont {Molnar}\ \emph {et~al.}(2018)\citenamefont {Molnar},
  \citenamefont {Ge}, \citenamefont {Schuch},\ and\ \citenamefont
  {Cirac}}]{Molnar2018}%
  \BibitemOpen
  \bibfield  {author} {\bibinfo {author} {\bibfnamefont {Andras}\ \bibnamefont
  {Molnar}}, \bibinfo {author} {\bibfnamefont {Yimin}\ \bibnamefont {Ge}},
  \bibinfo {author} {\bibfnamefont {Norbert}\ \bibnamefont {Schuch}}, \ and\
  \bibinfo {author} {\bibfnamefont {J~Ignacio}\ \bibnamefont {Cirac}},\
  }\bibfield  {title} {\enquote {\bibinfo {title} {{A generalization of the
  injectivity condition for projected entangled pair states}},}\ }\href
  {\doibase 10.1063/1.5007017} {\bibfield  {journal} {\bibinfo  {journal}
  {Journal of Mathematical Physics}\ }\textbf {\bibinfo {volume} {59}}
  (\bibinfo {year} {2018}),\ 10.1063/1.5007017}\BibitemShut {NoStop}%
\bibitem [{Note2()}]{Note2}%
  \BibitemOpen
  \bibinfo {note} {For all states we study in this paper, we simply set
  $\{Q_v\}$ on the boundary to be identity operators.}\BibitemShut {Stop}%
\bibitem [{\citenamefont {Perez-Garcia}\ \emph {et~al.}(2008)\citenamefont
  {Perez-Garcia}, \citenamefont {Verstraete}, \citenamefont {Wolf},\ and\
  \citenamefont {Cirac}}]{perez2008peps}%
  \BibitemOpen
  \bibfield  {author} {\bibinfo {author} {\bibfnamefont {David}\ \bibnamefont
  {Perez-Garcia}}, \bibinfo {author} {\bibfnamefont {Frank}\ \bibnamefont
  {Verstraete}}, \bibinfo {author} {\bibfnamefont {Michael~M}\ \bibnamefont
  {Wolf}}, \ and\ \bibinfo {author} {\bibfnamefont {J~Ignacio}\ \bibnamefont
  {Cirac}},\ }\bibfield  {title} {\enquote {\bibinfo {title} {{PEPS as unique
  ground states of local Hamiltonians}},}\ }\href
  {https://arxiv.org/abs/0707.2260} {\bibfield  {journal} {\bibinfo  {journal}
  {Quantum Information \& Computation}\ }\textbf {\bibinfo {volume} {8}},\
  \bibinfo {pages} {650--663} (\bibinfo {year} {2008})}\BibitemShut {NoStop}%
\bibitem [{Note3()}]{Note3}%
  \BibitemOpen
  \bibinfo {note} {The Hamiltonian [\protect \cref {tns_ham}] can be
  efficiently constructed for arbitrary 1D MPS and 2D translationally invariant
  (TI) PEPS, which covers all examples studied in this paper. However, for a
  generic non-TI 2D PEPS, one cannot scalably obtain the few-sites reduced
  density matrices $\{\rho _e\}$. In this case, it is possible to use a
  numerical optimization approach~\cite {Giudici2022rkq} to obtain a parent
  Hamiltonian with each term also only support on a few sites.}\BibitemShut
  {Stop}%
\bibitem [{Note4()}]{Note4}%
  \BibitemOpen
  \bibinfo {note} {One can transform $\{A_{g}\}$ to $\{A_{1/g}\}$ by first
  swapping the two indices in the auxiliary space (corresponding to transposing
  the matrices $\{A\}$), then swapping the physical index 2 and 3, and finally
  normalize the state.}\BibitemShut {Stop}%
\bibitem [{Note5()}]{Note5}%
  \BibitemOpen
  \bibinfo {note} {The initial state $\left |\psi (0) \right \rangle $ can be
  prepared with local quantum circuit or adiabatic evolution of a constant
  depth (time).}\BibitemShut {Stop}%
\bibitem [{Note6()}]{Note6}%
  \BibitemOpen
  \bibinfo {note} {We choose the interpolation functions such that they have a
  finite level of smoothness, which is both experimentally realistic and lead
  to an almost exponential decay of the error [c.f.~\protect \cref
  {cls_qpt_fid_scale}] in a wide range of evolution time $T$. Since in we need
  to reach larger system size $N\sim 5000$ in 1D compared to $N\sim 100$ in 2D,
  we choose the function $s(t,T)_{\protect \rm 1D}$ to be more smooth than
  $s(t,T)_{\protect \rm 2D}$ at $t=0$ and $t=T$.}\BibitemShut {Stop}%
\bibitem [{\citenamefont {Nenciu}(1993)}]{nenciu1993linear}%
  \BibitemOpen
  \bibfield  {author} {\bibinfo {author} {\bibfnamefont {Gheorghe}\
  \bibnamefont {Nenciu}},\ }\bibfield  {title} {\enquote {\bibinfo {title}
  {{Linear adiabatic theory. Exponential estimates}},}\ }\href
  {https://link.springer.com/article/10.1007/BF02096616} {\bibfield  {journal}
  {\bibinfo  {journal} {Communications in mathematical physics}\ }\textbf
  {\bibinfo {volume} {152}},\ \bibinfo {pages} {479--496} (\bibinfo {year}
  {1993})}\BibitemShut {NoStop}%
\bibitem [{\citenamefont {Hagedorn}\ and\ \citenamefont
  {Joye}(2002)}]{hagedorn2002elementary}%
  \BibitemOpen
  \bibfield  {author} {\bibinfo {author} {\bibfnamefont {George~A}\
  \bibnamefont {Hagedorn}}\ and\ \bibinfo {author} {\bibfnamefont {Alain}\
  \bibnamefont {Joye}},\ }\bibfield  {title} {\enquote {\bibinfo {title}
  {{Elementary exponential error estimates for the adiabatic approximation}},}\
  }\href {https://www.sciencedirect.com/science/article/pii/S0022247X01977650}
  {\bibfield  {journal} {\bibinfo  {journal} {Journal of mathematical analysis
  and applications}\ }\textbf {\bibinfo {volume} {267}},\ \bibinfo {pages}
  {235--246} (\bibinfo {year} {2002})}\BibitemShut {NoStop}%
\bibitem [{\citenamefont {Lidar}\ \emph {et~al.}(2009)\citenamefont {Lidar},
  \citenamefont {Rezakhani},\ and\ \citenamefont {Hamma}}]{lidar2009}%
  \BibitemOpen
  \bibfield  {author} {\bibinfo {author} {\bibfnamefont {Daniel~A.}\
  \bibnamefont {Lidar}}, \bibinfo {author} {\bibfnamefont {Ali~T.}\
  \bibnamefont {Rezakhani}}, \ and\ \bibinfo {author} {\bibfnamefont
  {Alioscia}\ \bibnamefont {Hamma}},\ }\bibfield  {title} {\enquote {\bibinfo
  {title} {{Adiabatic approximation with exponential accuracy for many-body
  systems and quantum computation}},}\ }\href {\doibase 10.1063/1.3236685}
  {\bibfield  {journal} {\bibinfo  {journal} {Journal of Mathematical Physics}\
  }\textbf {\bibinfo {volume} {50}} (\bibinfo {year} {2009}),\
  10.1063/1.3236685}\BibitemShut {NoStop}%
\bibitem [{\citenamefont {Rezakhani}\ \emph {et~al.}(2010)\citenamefont
  {Rezakhani}, \citenamefont {Pimachev},\ and\ \citenamefont
  {Lidar}}]{Rezakhani2010}%
  \BibitemOpen
  \bibfield  {author} {\bibinfo {author} {\bibfnamefont {A.~T.}\ \bibnamefont
  {Rezakhani}}, \bibinfo {author} {\bibfnamefont {A.~K.}\ \bibnamefont
  {Pimachev}}, \ and\ \bibinfo {author} {\bibfnamefont {D.~A.}\ \bibnamefont
  {Lidar}},\ }\bibfield  {title} {\enquote {\bibinfo {title} {{Accuracy versus
  run time in an adiabatic quantum search}},}\ }\href {\doibase
  10.1103/PhysRevA.82.052305} {\bibfield  {journal} {\bibinfo  {journal}
  {Physical Review A}\ }\textbf {\bibinfo {volume} {82}},\ \bibinfo {pages}
  {1--19} (\bibinfo {year} {2010})}\BibitemShut {NoStop}%
\bibitem [{Note7()}]{Note7}%
  \BibitemOpen
  \bibinfo {note} {Note that the sequential method prepares the state
  deterministically, while the adiabatic method always prepares the state
  approximately (with high fidelity), which is enough for practical purposes.
  Moreover, there could be a constant overhead to implement the Hamiltonian
  dynamics using quantum circuits, and this does not affect our statement that
  there will be a regime $(N\gg 1)$ where the adiabatic preparation will be
  more efficient.}\BibitemShut {Stop}%
\bibitem [{\citenamefont {Zhou}\ \emph {et~al.}(2021)\citenamefont {Zhou},
  \citenamefont {Choi},\ and\ \citenamefont {Lukin}}]{Zhou2021b}%
  \BibitemOpen
  \bibfield  {author} {\bibinfo {author} {\bibfnamefont {Leo}\ \bibnamefont
  {Zhou}}, \bibinfo {author} {\bibfnamefont {Soonwon}\ \bibnamefont {Choi}}, \
  and\ \bibinfo {author} {\bibfnamefont {Mikhail~D.}\ \bibnamefont {Lukin}},\
  }\bibfield  {title} {\enquote {\bibinfo {title} {{Symmetry-protected
  dissipative preparation of matrix product states}},}\ }\href {\doibase
  10.1103/PhysRevA.104.032418} {\bibfield  {journal} {\bibinfo  {journal}
  {Physical Review A}\ }\textbf {\bibinfo {volume} {104}},\ \bibinfo {pages}
  {1--19} (\bibinfo {year} {2021})}\BibitemShut {NoStop}%
\bibitem [{\citenamefont {Kaltenbaek}\ \emph {et~al.}(2010)\citenamefont
  {Kaltenbaek}, \citenamefont {Lavoie}, \citenamefont {Zeng}, \citenamefont
  {Bartlett},\ and\ \citenamefont {Resch}}]{Kaltenbaek2010}%
  \BibitemOpen
  \bibfield  {author} {\bibinfo {author} {\bibfnamefont {Rainer}\ \bibnamefont
  {Kaltenbaek}}, \bibinfo {author} {\bibfnamefont {Jonathan}\ \bibnamefont
  {Lavoie}}, \bibinfo {author} {\bibfnamefont {Bei}\ \bibnamefont {Zeng}},
  \bibinfo {author} {\bibfnamefont {Stephen~D.}\ \bibnamefont {Bartlett}}, \
  and\ \bibinfo {author} {\bibfnamefont {Kevin~J.}\ \bibnamefont {Resch}},\
  }\bibfield  {title} {\enquote {\bibinfo {title} {{Optical one-way quantum
  computing with a simulated valence-bond solid}},}\ }\href {\doibase
  10.1038/nphys1777} {\bibfield  {journal} {\bibinfo  {journal} {Nature
  Physics}\ }\textbf {\bibinfo {volume} {6}},\ \bibinfo {pages} {850--854}
  (\bibinfo {year} {2010})}\BibitemShut {NoStop}%
\bibitem [{\citenamefont {Lemm}\ \emph {et~al.}(2020)\citenamefont {Lemm},
  \citenamefont {Sandvik},\ and\ \citenamefont {Wang}}]{Lemm2020}%
  \BibitemOpen
  \bibfield  {author} {\bibinfo {author} {\bibfnamefont {Marius}\ \bibnamefont
  {Lemm}}, \bibinfo {author} {\bibfnamefont {Anders~W.}\ \bibnamefont
  {Sandvik}}, \ and\ \bibinfo {author} {\bibfnamefont {Ling}\ \bibnamefont
  {Wang}},\ }\bibfield  {title} {\enquote {\bibinfo {title} {{Existence of a
  Spectral Gap in the Affleck-Kennedy-Lieb-Tasaki Model on the Hexagonal
  Lattice}},}\ }\href {\doibase 10.1103/PhysRevLett.124.177204} {\bibfield
  {journal} {\bibinfo  {journal} {Physical Review Letters}\ }\textbf {\bibinfo
  {volume} {124}},\ \bibinfo {pages} {177204} (\bibinfo {year}
  {2020})}\BibitemShut {NoStop}%
\bibitem [{\citenamefont {Pomata}\ and\ \citenamefont
  {Wei}(2020)}]{Pomata2020}%
  \BibitemOpen
  \bibfield  {author} {\bibinfo {author} {\bibfnamefont {Nicholas}\
  \bibnamefont {Pomata}}\ and\ \bibinfo {author} {\bibfnamefont {Tzu~Chieh}\
  \bibnamefont {Wei}},\ }\bibfield  {title} {\enquote {\bibinfo {title}
  {{Demonstrating the Affleck-Kennedy-Lieb-Tasaki Spectral Gap on 2D Degree-3
  Lattices}},}\ }\href {\doibase 10.1103/PhysRevLett.124.177203} {\bibfield
  {journal} {\bibinfo  {journal} {Physical Review Letters}\ }\textbf {\bibinfo
  {volume} {124}},\ \bibinfo {pages} {177203} (\bibinfo {year}
  {2020})}\BibitemShut {NoStop}%
\bibitem [{\citenamefont {Koch-Janusz}\ \emph {et~al.}(2015)\citenamefont
  {Koch-Janusz}, \citenamefont {Khomskii},\ and\ \citenamefont
  {Sela}}]{koch2015affleck}%
  \BibitemOpen
  \bibfield  {author} {\bibinfo {author} {\bibfnamefont {Maciej}\ \bibnamefont
  {Koch-Janusz}}, \bibinfo {author} {\bibfnamefont {D~I}\ \bibnamefont
  {Khomskii}}, \ and\ \bibinfo {author} {\bibfnamefont {Eran}\ \bibnamefont
  {Sela}},\ }\bibfield  {title} {\enquote {\bibinfo {title}
  {{Affleck-Kennedy-Lieb-Tasaki State on a Honeycomb Lattice from t 2 g
  Orbitals}},}\ }\href
  {https://journals.aps.org/prl/abstract/10.1103/PhysRevLett.114.247204}
  {\bibfield  {journal} {\bibinfo  {journal} {Physical review letters}\
  }\textbf {\bibinfo {volume} {114}},\ \bibinfo {pages} {247204} (\bibinfo
  {year} {2015})}\BibitemShut {NoStop}%
\bibitem [{Note8()}]{Note8}%
  \BibitemOpen
  \bibinfo {note} {The 2D AKLT state studied here requires to block 6 sites in
  a hexagon [c.f.~\protect \cref {fig1}(c)] to make the tensor
  injective.}\BibitemShut {Stop}%
\bibitem [{\citenamefont {Lubasch}\ \emph {et~al.}(2014)\citenamefont
  {Lubasch}, \citenamefont {Cirac},\ and\ \citenamefont
  {Ba{\~{n}}uls}}]{Lubasch2014}%
  \BibitemOpen
  \bibfield  {author} {\bibinfo {author} {\bibfnamefont {Michael}\ \bibnamefont
  {Lubasch}}, \bibinfo {author} {\bibfnamefont {J.~Ignacio}\ \bibnamefont
  {Cirac}}, \ and\ \bibinfo {author} {\bibfnamefont {Mari~Carmen}\ \bibnamefont
  {Ba{\~{n}}uls}},\ }\bibfield  {title} {\enquote {\bibinfo {title}
  {{Algorithms for finite projected entangled pair states}},}\ }\href {\doibase
  10.1103/PhysRevB.90.064425} {\bibfield  {journal} {\bibinfo  {journal}
  {Physical Review B}\ }\textbf {\bibinfo {volume} {90}},\ \bibinfo {pages}
  {1--16} (\bibinfo {year} {2014})}\BibitemShut {NoStop}%
\bibitem [{Note9()}]{Note9}%
  \BibitemOpen
  \bibinfo {note} {This is also true for the path in Ref.~\cite
  {Ge2016,cruz2022preparation}. However, the large support of each Hamiltonian
  term makes it still difficult to simulate the adiabatic dynamics along that
  path.}\BibitemShut {Stop}%
\bibitem [{\citenamefont {Kastoryano}\ and\ \citenamefont
  {Lucia}(2018)}]{Kastoryano2018}%
  \BibitemOpen
  \bibfield  {author} {\bibinfo {author} {\bibfnamefont {Michael~J.}\
  \bibnamefont {Kastoryano}}\ and\ \bibinfo {author} {\bibfnamefont {Angelo}\
  \bibnamefont {Lucia}},\ }\bibfield  {title} {\enquote {\bibinfo {title}
  {{Divide and conquer method for proving gaps of frustration free
  Hamiltonians}},}\ }\href {\doibase 10.1088/1742-5468/aaa793} {\bibfield
  {journal} {\bibinfo  {journal} {Journal of Statistical Mechanics: Theory and
  Experiment}\ }\textbf {\bibinfo {volume} {2018}} (\bibinfo {year} {2018}),\
  10.1088/1742-5468/aaa793},\ \Eprint {http://arxiv.org/abs/1705.09491}
  {arXiv:1705.09491} \BibitemShut {NoStop}%
\bibitem [{\citenamefont {Mezzacapo}\ \emph {et~al.}(2014)\citenamefont
  {Mezzacapo}, \citenamefont {Lamata}, \citenamefont {Filipp},\ and\
  \citenamefont {Solano}}]{mezzacapo2014many}%
  \BibitemOpen
  \bibfield  {author} {\bibinfo {author} {\bibfnamefont {A}~\bibnamefont
  {Mezzacapo}}, \bibinfo {author} {\bibfnamefont {L}~\bibnamefont {Lamata}},
  \bibinfo {author} {\bibfnamefont {S}~\bibnamefont {Filipp}}, \ and\ \bibinfo
  {author} {\bibfnamefont {E}~\bibnamefont {Solano}},\ }\bibfield  {title}
  {\enquote {\bibinfo {title} {{Many-body interactions with tunable-coupling
  transmon qubits}},}\ }\href@noop {} {\bibfield  {journal} {\bibinfo
  {journal} {Physical review letters}\ }\textbf {\bibinfo {volume} {113}},\
  \bibinfo {pages} {50501} (\bibinfo {year} {2014})}\BibitemShut {NoStop}%
\bibitem [{\citenamefont {Bermudez}\ \emph {et~al.}(2009)\citenamefont
  {Bermudez}, \citenamefont {Porras},\ and\ \citenamefont
  {Martin-Delgado}}]{bermudez2009competing}%
  \BibitemOpen
  \bibfield  {author} {\bibinfo {author} {\bibfnamefont {A}~\bibnamefont
  {Bermudez}}, \bibinfo {author} {\bibfnamefont {Diego}\ \bibnamefont
  {Porras}}, \ and\ \bibinfo {author} {\bibfnamefont {M~A}\ \bibnamefont
  {Martin-Delgado}},\ }\bibfield  {title} {\enquote {\bibinfo {title}
  {{Competing many-body interactions in systems of trapped ions}},}\
  }\href@noop {} {\bibfield  {journal} {\bibinfo  {journal} {Physical review
  A}\ }\textbf {\bibinfo {volume} {79}},\ \bibinfo {pages} {60303} (\bibinfo
  {year} {2009})}\BibitemShut {NoStop}%
\bibitem [{\citenamefont {Scholl}\ \emph {et~al.}(2022)\citenamefont {Scholl},
  \citenamefont {Williams}, \citenamefont {Bornet}, \citenamefont {Wallner},
  \citenamefont {Barredo}, \citenamefont {Henriet}, \citenamefont {Signoles},
  \citenamefont {Hainaut}, \citenamefont {Franz}, \citenamefont {Geier},\ and\
  \citenamefont {Others}}]{scholl2022microwave}%
  \BibitemOpen
  \bibfield  {author} {\bibinfo {author} {\bibfnamefont {Pascal}\ \bibnamefont
  {Scholl}}, \bibinfo {author} {\bibfnamefont {Hannah~J}\ \bibnamefont
  {Williams}}, \bibinfo {author} {\bibfnamefont {Guillaume}\ \bibnamefont
  {Bornet}}, \bibinfo {author} {\bibfnamefont {Florian}\ \bibnamefont
  {Wallner}}, \bibinfo {author} {\bibfnamefont {Daniel}\ \bibnamefont
  {Barredo}}, \bibinfo {author} {\bibfnamefont {L}~\bibnamefont {Henriet}},
  \bibinfo {author} {\bibfnamefont {Adrien}\ \bibnamefont {Signoles}}, \bibinfo
  {author} {\bibfnamefont {Cl{\'{e}}ment}\ \bibnamefont {Hainaut}}, \bibinfo
  {author} {\bibfnamefont {Titus}\ \bibnamefont {Franz}}, \bibinfo {author}
  {\bibfnamefont {S}~\bibnamefont {Geier}}, \ and\ \bibinfo {author}
  {\bibnamefont {Others}},\ }\bibfield  {title} {\enquote {\bibinfo {title}
  {{Microwave Engineering of Programmable X X Z Hamiltonians in Arrays of
  Rydberg Atoms}},}\ }\href@noop {} {\bibfield  {journal} {\bibinfo  {journal}
  {PRX Quantum}\ }\textbf {\bibinfo {volume} {3}},\ \bibinfo {pages} {20303}
  (\bibinfo {year} {2022})}\BibitemShut {NoStop}%
\bibitem [{\citenamefont {Cubitt}\ \emph {et~al.}(2017)\citenamefont {Cubitt},
  \citenamefont {Montanaro},\ and\ \citenamefont {Piddock}}]{Cubitt2017}%
  \BibitemOpen
  \bibfield  {author} {\bibinfo {author} {\bibfnamefont {Toby}\ \bibnamefont
  {Cubitt}}, \bibinfo {author} {\bibfnamefont {Ashley}\ \bibnamefont
  {Montanaro}}, \ and\ \bibinfo {author} {\bibfnamefont {Stephen}\ \bibnamefont
  {Piddock}},\ }\bibfield  {title} {\enquote {\bibinfo {title} {{Universal
  Quantum Hamiltonians}},}\ }\href {\doibase 10.1073/pnas.1804949115} {\
  \textbf {\bibinfo {volume} {115}} (\bibinfo {year} {2017}),\
  10.1073/pnas.1804949115}\BibitemShut {NoStop}%
\bibitem [{\citenamefont {Zhou}\ and\ \citenamefont {Aharonov}()}]{Zhou2021a}%
  \BibitemOpen
  \bibfield  {author} {\bibinfo {author} {\bibfnamefont {Leo}\ \bibnamefont
  {Zhou}}\ and\ \bibinfo {author} {\bibfnamefont {Dorit}\ \bibnamefont
  {Aharonov}},\ }\bibfield  {title} {\enquote {\bibinfo {title} {{Strongly
  Universal Hamiltonian Simulators}},}\ }\href
  {http://arxiv.org/abs/2102.02991} {\ }\Eprint
  {http://arxiv.org/abs/2102.02991} {arXiv:2102.02991} \BibitemShut {NoStop}%
\bibitem [{\citenamefont {Choi}\ \emph {et~al.}(2020)\citenamefont {Choi},
  \citenamefont {Zhou}, \citenamefont {Knowles}, \citenamefont {Landig},
  \citenamefont {Choi},\ and\ \citenamefont {Lukin}}]{Choi2020}%
  \BibitemOpen
  \bibfield  {author} {\bibinfo {author} {\bibfnamefont {Joonhee}\ \bibnamefont
  {Choi}}, \bibinfo {author} {\bibfnamefont {Hengyun}\ \bibnamefont {Zhou}},
  \bibinfo {author} {\bibfnamefont {Helena~S.}\ \bibnamefont {Knowles}},
  \bibinfo {author} {\bibfnamefont {Renate}\ \bibnamefont {Landig}}, \bibinfo
  {author} {\bibfnamefont {Soonwon}\ \bibnamefont {Choi}}, \ and\ \bibinfo
  {author} {\bibfnamefont {Mikhail~D.}\ \bibnamefont {Lukin}},\ }\bibfield
  {title} {\enquote {\bibinfo {title} {{Robust Dynamic Hamiltonian Engineering
  of Many-Body Spin Systems}},}\ }\href {\doibase 10.1103/PHYSREVX.10.031002}
  {\bibfield  {journal} {\bibinfo  {journal} {Physical Review X}\ }\textbf
  {\bibinfo {volume} {10}},\ \bibinfo {pages} {31002} (\bibinfo {year}
  {2020})}\BibitemShut {NoStop}%
\bibitem [{\citenamefont {Fishman}\ \emph {et~al.}(2020)\citenamefont
  {Fishman}, \citenamefont {White},\ and\ \citenamefont
  {Stoudenmire}}]{itensor}%
  \BibitemOpen
  \bibfield  {author} {\bibinfo {author} {\bibfnamefont {Matthew}\ \bibnamefont
  {Fishman}}, \bibinfo {author} {\bibfnamefont {Steven~R}\ \bibnamefont
  {White}}, \ and\ \bibinfo {author} {\bibfnamefont {E~Miles}\ \bibnamefont
  {Stoudenmire}},\ }\href@noop {} {\enquote {\bibinfo {title} {{The
  \mbox{ITensor} Software Library for Tensor Network Calculations}},}\ }
  (\bibinfo {year} {2020}),\ \Eprint {http://arxiv.org/abs/2007.14822}
  {arXiv:2007.14822} \BibitemShut {NoStop}%
\bibitem [{\citenamefont {Murta}\ \emph {et~al.}(2022)\citenamefont {Murta},
  \citenamefont {Cruz},\ and\ \citenamefont
  {Fern{\'{a}}ndez-Rossier}}]{Murta2022}%
  \BibitemOpen
  \bibfield  {author} {\bibinfo {author} {\bibfnamefont {Bruno}\ \bibnamefont
  {Murta}}, \bibinfo {author} {\bibfnamefont {Pedro M~Q}\ \bibnamefont {Cruz}},
  \ and\ \bibinfo {author} {\bibfnamefont {J}~\bibnamefont
  {Fern{\'{a}}ndez-Rossier}},\ }\href {\doibase 10.48550/ARXIV.2207.07725}
  {\enquote {\bibinfo {title} {{Preparing Valence-Bond-Solid states on noisy
  intermediate-scale quantum computers}},}\ } (\bibinfo {year}
  {2022})\BibitemShut {NoStop}%
\bibitem [{\citenamefont {Giudici}\ \emph {et~al.}(2022)\citenamefont
  {Giudici}, \citenamefont {Cirac},\ and\ \citenamefont
  {Schuch}}]{Giudici2022rkq}%
  \BibitemOpen
  \bibfield  {author} {\bibinfo {author} {\bibfnamefont {Giuliano}\
  \bibnamefont {Giudici}}, \bibinfo {author} {\bibfnamefont {J~Ignacio}\
  \bibnamefont {Cirac}}, \ and\ \bibinfo {author} {\bibfnamefont {Norbert}\
  \bibnamefont {Schuch}},\ }\bibfield  {title} {\enquote {\bibinfo {title}
  {{Locality optimization for parent Hamiltonians of Tensor Networks}},}\
  }\href@noop {} {\  (\bibinfo {year} {2022})},\ \Eprint
  {http://arxiv.org/abs/2203.07443} {arXiv:2203.07443 [cond-mat.str-el]}
  \BibitemShut {NoStop}%
\end{thebibliography}%


%merlin.mbs apsrev4-1.bst 2010-07-25 4.21a (PWD, AO, DPC) hacked
%Control: key (0)
%Control: author (8) initials jnrlst
%Control: editor formatted (1) identically to author
%Control: production of article title (-1) disabled
%Control: page (0) single
%Control: year (1) truncated
%Control: production of eprint (0) enabled
\begin{thebibliography}{18}%
\makeatletter
\providecommand \@ifxundefined [1]{%
 \@ifx{#1\undefined}
}%
\providecommand \@ifnum [1]{%
 \ifnum #1\expandafter \@firstoftwo
 \else \expandafter \@secondoftwo
 \fi
}%
\providecommand \@ifx [1]{%
 \ifx #1\expandafter \@firstoftwo
 \else \expandafter \@secondoftwo
 \fi
}%
\providecommand \natexlab [1]{#1}%
\providecommand \enquote  [1]{``#1''}%
\providecommand \bibnamefont  [1]{#1}%
\providecommand \bibfnamefont [1]{#1}%
\providecommand \citenamefont [1]{#1}%
\providecommand \href@noop [0]{\@secondoftwo}%
\providecommand \href [0]{\begingroup \@sanitize@url \@href}%
\providecommand \@href[1]{\@@startlink{#1}\@@href}%
\providecommand \@@href[1]{\endgroup#1\@@endlink}%
\providecommand \@sanitize@url [0]{\catcode `\\12\catcode `\$12\catcode
  `\&12\catcode `\#12\catcode `\^12\catcode `\_12\catcode `\%12\relax}%
\providecommand \@@startlink[1]{}%
\providecommand \@@endlink[0]{}%
\providecommand \url  [0]{\begingroup\@sanitize@url \@url }%
\providecommand \@url [1]{\endgroup\@href {#1}{\urlprefix }}%
\providecommand \urlprefix  [0]{URL }%
\providecommand \Eprint [0]{\href }%
\providecommand \doibase [0]{http://dx.doi.org/}%
\providecommand \selectlanguage [0]{\@gobble}%
\providecommand \bibinfo  [0]{\@secondoftwo}%
\providecommand \bibfield  [0]{\@secondoftwo}%
\providecommand \translation [1]{[#1]}%
\providecommand \BibitemOpen [0]{}%
\providecommand \bibitemStop [0]{}%
\providecommand \bibitemNoStop [0]{.\EOS\space}%
\providecommand \EOS [0]{\spacefactor3000\relax}%
\providecommand \BibitemShut  [1]{\csname bibitem#1\endcsname}%
\let\auto@bib@innerbib\@empty
%</preamble>
\bibitem [{\citenamefont {Vidal}(2004)}]{Vidal2004}%
  \BibitemOpen
  \bibfield  {author} {\bibinfo {author} {\bibfnamefont {G.}~\bibnamefont
  {Vidal}},\ }\href {\doibase 10.1103/PhysRevLett.93.040502} {\bibfield
  {journal} {\bibinfo  {journal} {Physical Review Letters}\ }\textbf {\bibinfo
  {volume} {93}},\ \bibinfo {pages} {1} (\bibinfo {year} {2004})}\BibitemShut
  {NoStop}%
\bibitem [{\citenamefont {Hatano}\ and\ \citenamefont
  {Suzuki}(2005)}]{hatano2005finding}%
  \BibitemOpen
  \bibfield  {author} {\bibinfo {author} {\bibfnamefont {N.}~\bibnamefont
  {Hatano}}\ and\ \bibinfo {author} {\bibfnamefont {M.}~\bibnamefont
  {Suzuki}},\ }in\ \href {https://link.springer.com/chapter/10.1007/11526216_2}
  {\emph {\bibinfo {booktitle} {Quantum annealing and other optimization
  methods}}}\ (\bibinfo  {publisher} {Springer},\ \bibinfo {year} {2005})\ pp.\
  \bibinfo {pages} {37--68}\BibitemShut {NoStop}%
\bibitem [{\citenamefont {Fishman}\ \emph {et~al.}(2020)\citenamefont
  {Fishman}, \citenamefont {White},\ and\ \citenamefont
  {Stoudenmire}}]{itensor}%
  \BibitemOpen
  \bibfield  {author} {\bibinfo {author} {\bibfnamefont {M.}~\bibnamefont
  {Fishman}}, \bibinfo {author} {\bibfnamefont {S.~R.}\ \bibnamefont {White}},
  \ and\ \bibinfo {author} {\bibfnamefont {E.~M.}\ \bibnamefont
  {Stoudenmire}},\ }\href@noop {} {\enquote {\bibinfo {title} {{The
  \mbox{ITensor} Software Library for Tensor Network Calculations}},}\ }
  (\bibinfo {year} {2020}),\ \Eprint {http://arxiv.org/abs/2007.14822}
  {arXiv:2007.14822} \BibitemShut {NoStop}%
\bibitem [{\citenamefont {Schollw{\"{o}}ck}(2011)}]{schollwock2011density}%
  \BibitemOpen
  \bibfield  {author} {\bibinfo {author} {\bibfnamefont {U.}~\bibnamefont
  {Schollw{\"{o}}ck}},\ }\href
  {https://www.sciencedirect.com/science/article/abs/pii/S0003491610001752}
  {\bibfield  {journal} {\bibinfo  {journal} {Annals of Physics}\ }\textbf
  {\bibinfo {volume} {326}},\ \bibinfo {pages} {96} (\bibinfo {year}
  {2011})}\BibitemShut {NoStop}%
\bibitem [{\citenamefont {Nenciu}(1993)}]{nenciu1993linear}%
  \BibitemOpen
  \bibfield  {author} {\bibinfo {author} {\bibfnamefont {G.}~\bibnamefont
  {Nenciu}},\ }\href {https://link.springer.com/article/10.1007/BF02096616}
  {\bibfield  {journal} {\bibinfo  {journal} {Communications in mathematical
  physics}\ }\textbf {\bibinfo {volume} {152}},\ \bibinfo {pages} {479}
  (\bibinfo {year} {1993})}\BibitemShut {NoStop}%
\bibitem [{\citenamefont {Hagedorn}\ and\ \citenamefont
  {Joye}(2002)}]{hagedorn2002elementary}%
  \BibitemOpen
  \bibfield  {author} {\bibinfo {author} {\bibfnamefont {G.~A.}\ \bibnamefont
  {Hagedorn}}\ and\ \bibinfo {author} {\bibfnamefont {A.}~\bibnamefont
  {Joye}},\ }\href
  {https://www.sciencedirect.com/science/article/pii/S0022247X01977650}
  {\bibfield  {journal} {\bibinfo  {journal} {Journal of mathematical analysis
  and applications}\ }\textbf {\bibinfo {volume} {267}},\ \bibinfo {pages}
  {235} (\bibinfo {year} {2002})}\BibitemShut {NoStop}%
\bibitem [{\citenamefont {Lidar}\ \emph {et~al.}(2009)\citenamefont {Lidar},
  \citenamefont {Rezakhani},\ and\ \citenamefont {Hamma}}]{lidar2009}%
  \BibitemOpen
  \bibfield  {author} {\bibinfo {author} {\bibfnamefont {D.~A.}\ \bibnamefont
  {Lidar}}, \bibinfo {author} {\bibfnamefont {A.~T.}\ \bibnamefont
  {Rezakhani}}, \ and\ \bibinfo {author} {\bibfnamefont {A.}~\bibnamefont
  {Hamma}},\ }\href {\doibase 10.1063/1.3236685} {\bibfield  {journal}
  {\bibinfo  {journal} {Journal of Mathematical Physics}\ }\textbf {\bibinfo
  {volume} {50}} (\bibinfo {year} {2009}),\ 10.1063/1.3236685}\BibitemShut
  {NoStop}%
\bibitem [{\citenamefont {Rezakhani}\ \emph {et~al.}(2010)\citenamefont
  {Rezakhani}, \citenamefont {Pimachev},\ and\ \citenamefont
  {Lidar}}]{Rezakhani2010}%
  \BibitemOpen
  \bibfield  {author} {\bibinfo {author} {\bibfnamefont {A.~T.}\ \bibnamefont
  {Rezakhani}}, \bibinfo {author} {\bibfnamefont {A.~K.}\ \bibnamefont
  {Pimachev}}, \ and\ \bibinfo {author} {\bibfnamefont {D.~A.}\ \bibnamefont
  {Lidar}},\ }\href {\doibase 10.1103/PhysRevA.82.052305} {\bibfield  {journal}
  {\bibinfo  {journal} {Physical Review A}\ }\textbf {\bibinfo {volume} {82}},\
  \bibinfo {pages} {1} (\bibinfo {year} {2010})}\BibitemShut {NoStop}%
\bibitem [{\citenamefont {Gevrey}(1918)}]{gevrey1918nature}%
  \BibitemOpen
  \bibfield  {author} {\bibinfo {author} {\bibfnamefont {M.}~\bibnamefont
  {Gevrey}},\ }in\ \href {http://www.numdam.org/item/10.24033/asens.706.pdf}
  {\emph {\bibinfo {booktitle} {Annales Scientifiques de l'Ecole Normale
  Superieure}}},\ Vol.~\bibinfo {volume} {35}\ (\bibinfo {year} {1918})\ pp.\
  \bibinfo {pages} {129--190}\BibitemShut {NoStop}%
\bibitem [{\citenamefont {Ge}\ \emph {et~al.}(2016)\citenamefont {Ge},
  \citenamefont {Moln{\'{a}}r},\ and\ \citenamefont {Cirac}}]{Ge2016}%
  \BibitemOpen
  \bibfield  {author} {\bibinfo {author} {\bibfnamefont {Y.}~\bibnamefont
  {Ge}}, \bibinfo {author} {\bibfnamefont {A.}~\bibnamefont {Moln{\'{a}}r}}, \
  and\ \bibinfo {author} {\bibfnamefont {J.~I.}\ \bibnamefont {Cirac}},\ }\href
  {\doibase 10.1103/PhysRevLett.116.080503} {\bibfield  {journal} {\bibinfo
  {journal} {Physical Review Letters}\ }\textbf {\bibinfo {volume} {116}},\
  \bibinfo {pages} {1} (\bibinfo {year} {2016})}\BibitemShut {NoStop}%
\bibitem [{\citenamefont {Cirac}\ \emph {et~al.}(2021)\citenamefont {Cirac},
  \citenamefont {Perez-Garcia}, \citenamefont {Schuch},\ and\ \citenamefont
  {Verstraete}}]{cirac2021matrix}%
  \BibitemOpen
  \bibfield  {author} {\bibinfo {author} {\bibfnamefont {J.~I.}\ \bibnamefont
  {Cirac}}, \bibinfo {author} {\bibfnamefont {D.}~\bibnamefont {Perez-Garcia}},
  \bibinfo {author} {\bibfnamefont {N.}~\bibnamefont {Schuch}}, \ and\ \bibinfo
  {author} {\bibfnamefont {F.}~\bibnamefont {Verstraete}},\ }\href
  {https://journals.aps.org/rmp/abstract/10.1103/RevModPhys.93.045003}
  {\bibfield  {journal} {\bibinfo  {journal} {Reviews of Modern Physics}\
  }\textbf {\bibinfo {volume} {93}},\ \bibinfo {pages} {45003} (\bibinfo {year}
  {2021})}\BibitemShut {NoStop}%
\bibitem [{\citenamefont {Affleck}\ \emph {et~al.}(1987)\citenamefont
  {Affleck}, \citenamefont {Kennedy}, \citenamefont {Lieb},\ and\ \citenamefont
  {Tasaki}}]{Affleck1987}%
  \BibitemOpen
  \bibfield  {author} {\bibinfo {author} {\bibfnamefont {I.}~\bibnamefont
  {Affleck}}, \bibinfo {author} {\bibfnamefont {T.}~\bibnamefont {Kennedy}},
  \bibinfo {author} {\bibfnamefont {E.~H.}\ \bibnamefont {Lieb}}, \ and\
  \bibinfo {author} {\bibfnamefont {H.}~\bibnamefont {Tasaki}},\ }\href
  {\doibase 10.1103/PhysRevLett.59.799} {\bibfield  {journal} {\bibinfo
  {journal} {Physical Review Letters}\ }\textbf {\bibinfo {volume} {59}},\
  \bibinfo {pages} {799} (\bibinfo {year} {1987})}\BibitemShut {NoStop}%
\bibitem [{\citenamefont {Affleck}\ \emph {et~al.}(1988)\citenamefont
  {Affleck}, \citenamefont {Kennedy}, \citenamefont {Lieb},\ and\ \citenamefont
  {Tasaki}}]{Affleck1988}%
  \BibitemOpen
  \bibfield  {author} {\bibinfo {author} {\bibfnamefont {I.}~\bibnamefont
  {Affleck}}, \bibinfo {author} {\bibfnamefont {T.}~\bibnamefont {Kennedy}},
  \bibinfo {author} {\bibfnamefont {E.~H.}\ \bibnamefont {Lieb}}, \ and\
  \bibinfo {author} {\bibfnamefont {H.}~\bibnamefont {Tasaki}},\ }\href
  {\doibase 10.1007/BF01218021} {\bibfield  {journal} {\bibinfo  {journal}
  {Communications in Mathematical Physics}\ }\textbf {\bibinfo {volume}
  {115}},\ \bibinfo {pages} {477} (\bibinfo {year} {1988})}\BibitemShut
  {NoStop}%
\bibitem [{\citenamefont {Haegeman}\ \emph {et~al.}(2011)\citenamefont
  {Haegeman}, \citenamefont {Cirac}, \citenamefont {Osborne}, \citenamefont
  {Pizorn}, \citenamefont {Verschelde},\ and\ \citenamefont
  {Verstraete}}]{Haegeman2011}%
  \BibitemOpen
  \bibfield  {author} {\bibinfo {author} {\bibfnamefont {J.}~\bibnamefont
  {Haegeman}}, \bibinfo {author} {\bibfnamefont {J.~I.}\ \bibnamefont {Cirac}},
  \bibinfo {author} {\bibfnamefont {T.~J.}\ \bibnamefont {Osborne}}, \bibinfo
  {author} {\bibfnamefont {I.}~\bibnamefont {Pizorn}}, \bibinfo {author}
  {\bibfnamefont {H.}~\bibnamefont {Verschelde}}, \ and\ \bibinfo {author}
  {\bibfnamefont {F.}~\bibnamefont {Verstraete}},\ }\href {\doibase
  10.1103/PhysRevLett.107.070601} {\bibfield  {journal} {\bibinfo  {journal}
  {Physical Review Letters}\ }\textbf {\bibinfo {volume} {107}},\ \bibinfo
  {pages} {1} (\bibinfo {year} {2011})}\BibitemShut {NoStop}%
\bibitem [{\citenamefont {Haegeman}\ \emph {et~al.}(2016)\citenamefont
  {Haegeman}, \citenamefont {Lubich}, \citenamefont {Oseledets}, \citenamefont
  {Vandereycken},\ and\ \citenamefont {Verstraete}}]{Haegeman2016}%
  \BibitemOpen
  \bibfield  {author} {\bibinfo {author} {\bibfnamefont {J.}~\bibnamefont
  {Haegeman}}, \bibinfo {author} {\bibfnamefont {C.}~\bibnamefont {Lubich}},
  \bibinfo {author} {\bibfnamefont {I.}~\bibnamefont {Oseledets}}, \bibinfo
  {author} {\bibfnamefont {B.}~\bibnamefont {Vandereycken}}, \ and\ \bibinfo
  {author} {\bibfnamefont {F.}~\bibnamefont {Verstraete}},\ }\href {\doibase
  10.1103/PhysRevB.94.165116} {\bibfield  {journal} {\bibinfo  {journal}
  {Physical Review B}\ }\textbf {\bibinfo {volume} {94}},\ \bibinfo {pages} {1}
  (\bibinfo {year} {2016})}\BibitemShut {NoStop}%
\bibitem [{\citenamefont {Paeckel}\ \emph {et~al.}(2019)\citenamefont
  {Paeckel}, \citenamefont {K{\"{o}}hler}, \citenamefont {Swoboda},
  \citenamefont {Manmana}, \citenamefont {Schollw{\"{o}}ck},\ and\
  \citenamefont {Hubig}}]{paeckel2019time}%
  \BibitemOpen
  \bibfield  {author} {\bibinfo {author} {\bibfnamefont {S.}~\bibnamefont
  {Paeckel}}, \bibinfo {author} {\bibfnamefont {T.}~\bibnamefont
  {K{\"{o}}hler}}, \bibinfo {author} {\bibfnamefont {A.}~\bibnamefont
  {Swoboda}}, \bibinfo {author} {\bibfnamefont {S.~R.}\ \bibnamefont
  {Manmana}}, \bibinfo {author} {\bibfnamefont {U.}~\bibnamefont
  {Schollw{\"{o}}ck}}, \ and\ \bibinfo {author} {\bibfnamefont
  {C.}~\bibnamefont {Hubig}},\ }\href
  {https://www.sciencedirect.com/science/article/pii/S0003491619302532}
  {\bibfield  {journal} {\bibinfo  {journal} {Annals of Physics}\ }\textbf
  {\bibinfo {volume} {411}},\ \bibinfo {pages} {167998} (\bibinfo {year}
  {2019})}\BibitemShut {NoStop}%
\bibitem [{\citenamefont {Cruz}\ \emph {et~al.}(2022)\citenamefont {Cruz},
  \citenamefont {Baccari}, \citenamefont {Tura}, \citenamefont {Schuch},\ and\
  \citenamefont {Cirac}}]{cruz2022preparation}%
  \BibitemOpen
  \bibfield  {author} {\bibinfo {author} {\bibfnamefont {E.}~\bibnamefont
  {Cruz}}, \bibinfo {author} {\bibfnamefont {F.}~\bibnamefont {Baccari}},
  \bibinfo {author} {\bibfnamefont {J.}~\bibnamefont {Tura}}, \bibinfo {author}
  {\bibfnamefont {N.}~\bibnamefont {Schuch}}, \ and\ \bibinfo {author}
  {\bibfnamefont {J.~I.}\ \bibnamefont {Cirac}},\ }\href
  {https://journals.aps.org/prresearch/abstract/10.1103/PhysRevResearch.4.023161}
  {\bibfield  {journal} {\bibinfo  {journal} {Physical Review Research}\
  }\textbf {\bibinfo {volume} {4}},\ \bibinfo {pages} {23161} (\bibinfo {year}
  {2022})}\BibitemShut {NoStop}%
\bibitem [{\citenamefont {Molnar}\ \emph {et~al.}(2018)\citenamefont {Molnar},
  \citenamefont {Ge}, \citenamefont {Schuch},\ and\ \citenamefont
  {Cirac}}]{Molnar2018}%
  \BibitemOpen
  \bibfield  {author} {\bibinfo {author} {\bibfnamefont {A.}~\bibnamefont
  {Molnar}}, \bibinfo {author} {\bibfnamefont {Y.}~\bibnamefont {Ge}}, \bibinfo
  {author} {\bibfnamefont {N.}~\bibnamefont {Schuch}}, \ and\ \bibinfo {author}
  {\bibfnamefont {J.~I.}\ \bibnamefont {Cirac}},\ }\href {\doibase
  10.1063/1.5007017} {\bibfield  {journal} {\bibinfo  {journal} {Journal of
  Mathematical Physics}\ }\textbf {\bibinfo {volume} {59}} (\bibinfo {year}
  {2018}),\ 10.1063/1.5007017}\BibitemShut {NoStop}%
\end{thebibliography}%

\end{document}

\crefname{equation}{Eq.}{Eqs.}
\crefname{figure}{Fig.}{Fig.}
\crefname{appendix}{Appendix}{Appendix}
\renewcommand{\thefigure}{S\arabic{figure}}
\renewcommand{\theequation}{S\arabic{equation}}
\renewcommand{\bibnumfmt}[1]{[S#1]}
\renewcommand{\citenumfont}[1]{S#1}
\title{Supplementary Materials for: \\ Efficient Adiabatic Preparation of Tensor Network States}
\author{Zhi-Yuan Wei}
\author{Daniel Malz}
\author{J. Ignacio Cirac}
\affiliation{Max-Planck-Institut f{\"u}r Quantenoptik, Hans-Kopfermann-Stra{\ss}e 1, D-85748 Garching, Germany}
\affiliation{Munich Center for Quantum Science and Technology (MCQST), Schellingstr. 4, D-80799 M{\"u}nchen, Germany}
\date{\today}
\maketitle

\onecolumngrid
\section{Obtaining positive-semidefinite operators $\{Q_v\}$ for PEPS}
\label{posQ_construct}
For an arbitrary PEPS [Eq.(1) in the main text], each operator $Q_v$ on site $v\in {\cal V}$ maps $n_v$ virtual qub(d)its of $D$-level into a $d$-level physical site. One can use the polar decomposition to decompose each operator $Q_v$ as ${Q_v } = {U_v }{Q_v' }$, where ${U_v} $ is an isometry of dimension $d\times D^{n_v}$ and ${Q_v' }$ is a positive-semidefinite operator of dimension $D^{n_v}\times D^{n_v}$. Thus one can rewrite the Eq.(1) in the main text as 
\begin{equation} \label{psi_f_targ_up}
		\left|\psi\right\rangle \propto{ \bigotimes_{v \in \cal V}  U_{v}  \bigotimes_{v \in \cal V} } Q'_{v} {\bigotimes_{e \in \cal E} \left| {{\Phi ^ +  }} \right\rangle _e }.
\end{equation}
Thus up to a single layer of local isometries $\bigotimes_{v \in \cal V}  U_{v}$, one can take the operators $\{Q_v\}$ to be positive-semidefinite. Note that, to physically implement the isometries, we require $d\geq D^{n_v}$. This can be realized by blocking neighboring sites, or promoting the virtual qub(d)its on each site into physical ones in the case of AKLT states.

Therefore, during the preparation of the target PEPS with operators $\{Q_v \}$, first, we use the adiabatic path to prepare the state with operators $\{Q_v' \}$, then apply a single layer of local isometries $\{U_v \}$ at the end of the adiabatic preparation. As it takes a constant time $T_{\rm iso} = O(1)$ to implement the isometries, in the main text we focus on the adiabatic evolution time $T$, and the total state preparation time is $T_{\rm tot} = T + T_{\rm iso}$.

Note that, for AKLT states, one does not need to implement local isometries after the adiabatic preparation (i.e. $T_{\rm iso}=0$). Thus the adiabatic preparation time $T$ shown in Fig.2(b) and Fig.3 in the main text for AKLT states are the total preparation time of the state.

\section{Details on the adiabatic preparation of MPS}
\label{apd_err_den}

\subsection{1D TEBD simulation}
\label{tebd_simu}
We use the TEBD algorithm~\cite{Vidal2004} to study the one-dimensional adiabatic dynamics. 
Given a local Hamiltonian $H(t) = \sum\nolimits_{i  = 1}^{{N_p}} {{h_i }\left( t \right)} $ with $t=n\tau$, we approximate each step ${U_{t} } = {e^{ - iH\left( t \right)\tau }}$ by following Trotter-Suzuki approximation~\cite{hatano2005finding}
\begin{equation} \label{}
{U_{t} } \approx {e^{ - i{h_1(t)}/2}}{e^{ - i{h_2}(t)\tau /2}} \cdots {e^{ - i{h_{{N_p}}(t)}\tau /2}}{e^{ - i{h_{{N_p}}(t)}\tau /2}}{e^{ - i{h_{{N_p} - 1}}(t) \tau /2}} \cdots {e^{ - i{h_1}(t)\tau /2}} + O\left( {{\tau ^3}} \right),
\end{equation}
and apply such unitaries to MPS to evolve the state. The precision of this algorithm can be tuned by changing the Trotter step $\tau$ and a cutoff parameter $\delta$ (in ITensor~\cite{itensor}) that controls the precision of applying a unitary to MPS. We typically observe the convergence of results with a Trotter time step $\tau  = 0.04$ and $\delta=1\times10^{-10}$ for the data shown in this paper.

Note we are able to use TEBD because the state during the adiabatic evolution [c.f.~Eq.(6) in the main text] remains close to the ground state of the parent Hamiltonian, which is an MPS of bond dimension $D=2$.

\subsection{Additional numerical results}

\begin{figure}[h!]
	\centering
	\includegraphics[width=0.8\textwidth]{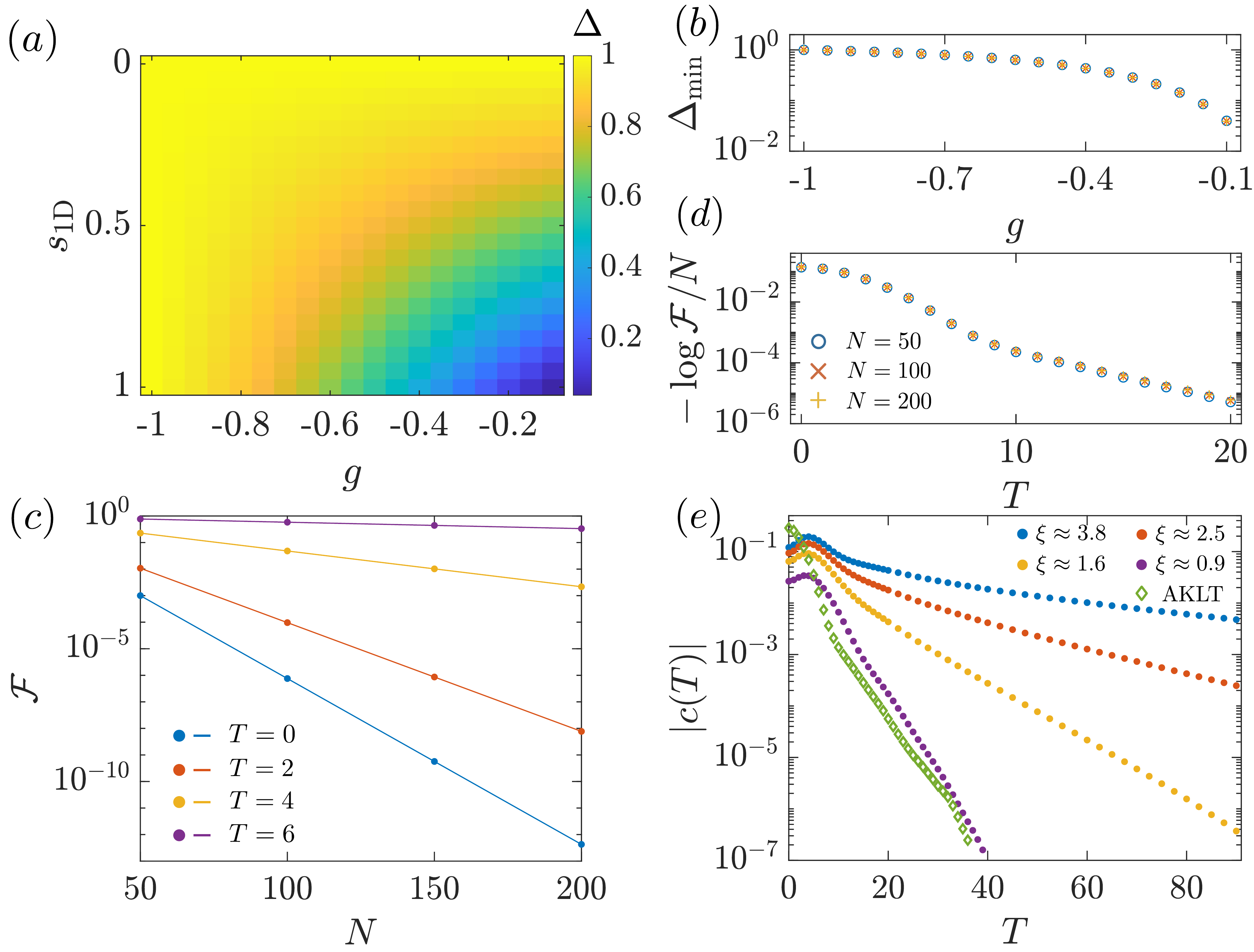}
        \caption{ (a) The gap spectroscopy of the adiabatic path [c.f.~Eq.(6) in the main text] for states in the MPS family with various $g$, computed with system size $N=100$. (b) The minimal gap $\Delta_{\rm min}$ during the adiabatic path [c.f.~Eq.(6) in the main text] for the MPS family with various $g$, extracted from the gap spectroscopy like that in (a) with various system size $N$. (c) The fidelity $\cal F$ of preparing the 1D AKLT state of different system size $N$, with various adiabatic evolution time $T$. (d) $- \log {\cal F}/N$ for preparing the 1D AKLT state as a function of the adiabatic evolution time $T$ for various system size $N$. (e) The evolution of the boundary term $c(T)$ [c.f.~Eq.(7) in the main text] for various states in the MPS family and the 1D AKLT state.}
        \label{exp_scale}
\end{figure}

In Fig.1(b) in the main text, we show the minimal gap $\Delta_{\rm min}$ during the adiabatic path for preparing states in the MPS family with different $g$. This is obtained using the density-matrix renormalization group (DMRG) method~\cite{schollwock2011density}. We present more data on the gap in \cref{exp_scale}(a). In \cref{exp_scale}(b) we show that the gap $\Delta_{\rm min}$ computed with different system sizes does not change with $N$.

\cref{exp_scale}(c) demonstrates the way we obtain the error density $\kappa(T)$ and the boundary term $c(T)$ in Eq.(7) in the main text. For each evolution time $T$, we simulate the quasi-adiabatic evolution of time $T$ for preparing the given MPS of various fixed system sizes $N=50,100,150,200$, and observe that the fidelity ${\cal F}(N,T)$ decay exponentially with $N$ for all fixed $T$. Thus we fit (numerically obtained) ${\cal F}(N,T)$ using Eq.(7) in the main text to extract $\kappa(T)$ and $c(T)$. In \cref{exp_scale}(d), we plot $-\log {\cal F}/N$ as a function of evolution time $T$ for various $N$. According to Eq.(7) in the main text, one expects
\begin{equation} \label{}
-\log {\cal F}(N,T)/N = \kappa(T)+c(T)/N \approx \kappa(T),
\end{equation}
and we see the collapse of data of different $N$ for all time $T$. Therefore, \cref{exp_scale}(c,d) demonstrate the correctness of Eq.(7) in the main text.

In \cref{exp_scale}(e) we plot $c(T)$ for various states in the MPS family and the AKLT state. As $c(T)$ represents errors on the boundary of the chain, it also shows an exponential decay with $T$ (thus $|c(T)|\rightarrow 0$), similar to the behavior of $\kappa(T)$ in Fig.2(a) in the main text.

\section{Asymptotic power-law decay of the error density}
\label{exp_to_pow}

For the MPS preparation in the main text, we use simple interpolation functions to interpolate between the initial Hamiltonian $H(0)$ and the final Hamiltonian $H(1)$ [c.f.~Eq.(3,6) in the main text], which leads to the good performance of the adiabatic algorithm shown in Fig.2 in the main text. However, as $s(t/T)$ we use is not \textit{infinitely} smooth, in the long-time limit, the error density $\kappa(T)$ cannot decay exponentially [c.f.~Eq.(8) in the main text], but rather scale as a power-law of the adiabatic evolution time $T$~\cite{nenciu1993linear,hagedorn2002elementary,lidar2009,Rezakhani2010}.

To understand this behavior, here we employ an exact diagonalization (ED) method to study the evolution of the same adiabatic dynamics [c.f.~Eq.(6) in the main text] with a small system size $N=8$, which can enter very small error regimes that $\kappa(T)\ll 1$. Thanks to the scaling between fidelity $\cal F$ and $N$ [c.f. Eq.(7) in the main text], the error density obtained for such a small system size could already provide a qualitative understanding of the behavior that also applies to larger systems.

Previous work has shown that, for a few-body system, the location where the behavior of $\kappa(T)$ transition from the exponential decay to power-law decay is mainly determined by the \textit{smoothness} of $s(t/T)$ at the beginning ($t=0$) and end ($t=T$) of the adiabatic evolution~\cite{Rezakhani2010}. So here we use the incomplete Beta functions $s(t/T)= \theta_k(t/T)$ in the adiabatic path [c.f.~Eq.(6) in the main text], which are given through
	\begin{equation} \label{inc_bt}
	\theta_{k}(\lambda)=\frac{B_{\lambda}(1+k, 1+k)}{B_{1}(1+k, 1+k)},
\end{equation}
with $B_{\lambda}(a, b) \equiv \int_{0}^{\lambda} y^{a-1}(1-y)^{b-1} d y$.
This function satisfies $k$-th order smoothness at $\lambda=0$ and $\lambda=1$, namely 
\begin{equation} \label{beta_func}
	\left.\frac{d^{n}\theta_{k}}{d\lambda^{n}}\right|_{\lambda=0,1}=0, \quad \forall n<k.
\end{equation}
Thus by controlling the order $k$, one can systematically control the smoothness of the interpolation function.

\begin{figure}[h!]
	\centering
	\includegraphics[width=0.8\textwidth]{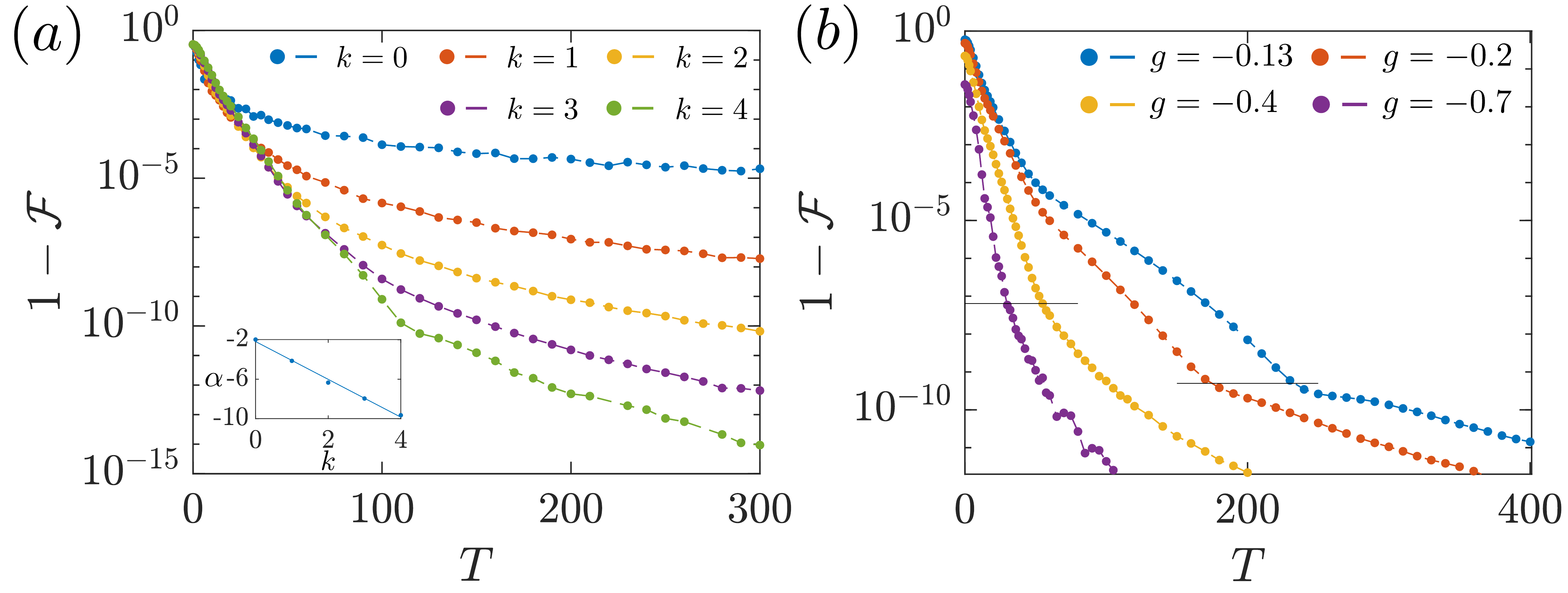}
        \caption{(a) The evolution of the infidelity $1-\cal F$ as a function of total adiabatic evolution time $T$ for preparing the MPS family with $g=-0.3$ of a system size $N=8$ using interpolation functions [c.f.~\cref{inc_bt}] with various order of smoothness $k$. The dashed lines are visual guides. The inset shows the power of the scaling \cref{power_err} as a function of the order of smoothness $k$, obtained by fitting the late-time evolution of $1-\cal F$ in (a). (b) The evolution of the infidelity for preparing the MPS family with various $g$, with a fixed order of smoothness $k=3$. The colored lines are visual guides, and the horizontal line denotes the infidelity level where the transition from exponential to power-law behavior approximately happens.}
        \label{fig_ed}
\end{figure}

\cref{fig_ed}(a) show the evolution of the infidelity $1-\cal F$ for preparing the MPS family with $g=-0.3$ using adiabatic evolution with different interpolation functions [c.f.~\cref{beta_func}]. One can see an exponential decay of the infidelity in a short time and then transition to a power-law decay regime at a late time. By increasing $k$, the exponential decay region can be increased, and the late-time power-law decay behaviors are
\begin{equation} \label{power_err}
	\kappa({T \gg 1}) \sim O(1/T^{\alpha}),
\end{equation}
with $\alpha \approx 2k+2$ shown in the inset of \cref{fig_ed}(a). This behavior agrees with that in Ref.~\cite{lidar2009,Rezakhani2010}. Also in \cref{fig_ed}(b), we show the evolution of the infidelity for preparing the MPS family with various $g$ using a fixed interpolation function ($k=3$) [c.f.~\cref{beta_func}]. For states with different correlation lengths, we see the transition from the exponential decay regime to the power-law decay regime generally happens when the infidelity is very small. Therefore, by choosing a $s(t/T)$ that is smooth enough, we expect the exponential decay of the error density $\kappa(T)$ can last to practically relevant (and potentially very large) system sizes. 

Note that, the above analysis does not mean that we should choose a too smooth $s(t/T)$. As Ref.~\cite{Rezakhani2010} has shown, the exponential decay rate $\gamma$ [c.f.~Eq.(8) in the main text] decreases with the smoothness of $s(t/T)$. Moreover, when choosing $s(t/T)$ to be \textit{infinitely} smooth at $t=0$ and $t=T$, it has been shown for the case of the Gevrey class $1+\alpha$ with $\alpha>0$~\cite{gevrey1918nature} that the error density $\kappa_{\rm Gev}(T)$ [c.f.~Eq.(7) in the main text] evolves as~\cite{Ge2016}
\begin{equation} \label{}
	\kappa_{\rm Gev}(T) \sim O\left[\exp(-{C(\alpha)} \cdot T^{\frac{1}{1+a}})\right],
\end{equation}
which is a slower-than-exponential decay since $\alpha > 0$. Therefore, in practice, one needs to choose the level of smoothness of $s(t/T)$ appropriately to minimize the overall infidelity.

\section{Details on the adiabatic preparation of AKLT states}

\subsection{Projectors of the AKLT states}
\label{apd_ak2d_hex}

Both the 1D AKLT state of spin $S=1$ and the 2D AKLT state of spin $S=3/2$ on the hexagonal lattice can be written as PEPS of the form [Eq.(1) in the main text] of bond dimension $D=2$. Here each operators $Q_v$ in the bulk consist of a projector $P_{S,v}$ and singlet matrix $Y=\left(\begin{array}{cc}
0 & -1 \\
1 & 0
\end{array}\right)$, where $P_{S,v}$ project the $n_v$ qubits on the vertex $v$ to their symmetric subspace, and each $Y$ converts an entangled pair into a singlet~\cite{cirac2021matrix}. For 1D, 
$Q^{1D}_v = P_{1,v} ({\mathbb 1} \bigotimes Y)$,
with
\begin{equation} \label{}
	P_{1,v} =  | 0,0 \rangle \langle 0,0 | + | 1,1 \rangle \langle 1,1 | 
	+ \frac{1}{2} (| 0,1 \rangle  + | 1,0 \rangle )(\langle 0,1 |  + \langle 1,0 | ).
\end{equation}
For 2D hexagonal lattice, $Q_v = P_{3/2,v} ({\mathbb 1} \bigotimes Y \bigotimes Y)$, with
\begin{align} \label{}
	P_{3/2,v} &=  | 0,0,0 \rangle \langle 0,0,0 | + | 1,1,1 \rangle \langle 1,1,1 | \\
	&+ \frac{1}{3} (| 0,1,1 \rangle  + | 1,0,1 \rangle + | 1,1,0 \rangle)(\langle 0,1,1 |  + \langle 1,0,1 | + \langle 1,1,0 |) \nonumber\\
	&+ \frac{1}{3} (| 0,0,1 \rangle  + | 0,1,0 \rangle + | 1,0,0 \rangle)(\langle 0,0,1 |  + \langle 0,1,0 | + \langle 1,0,0 |). \nonumber
\end{align}

Note that, here we have promoted the virtual qubits in AKLT states into physical ones, i.e. for the 1D (2D) case, there are $n_v=2$ ($n_v=3$) qubits on each site. The $S=1$ ($S=3/2$) spin in the original definition corresponds to the symmetric subspace of each local site here, with the basis~\cite{Affleck1987, Affleck1988}
\begin{equation} \label{}
S=1: \qquad	| S_z= 1 \rangle = | 0,0 \rangle, \qquad | S_z= 0 \rangle = \frac{1}{\sqrt{2}}(| 0,1 \rangle + | 1,0 \rangle), \qquad | S_z= -1 \rangle = | 1,1 \rangle,
\end{equation}
\begin{align}
S=3/2: \qquad	| S_z= 3/2 \rangle &= | 0,0,0 \rangle, \qquad | S_z= 1/2 \rangle = \frac{1}{\sqrt{3}}(| 0,0,1 \rangle + | 0,1,0 \rangle + | 1,0,0 \rangle), \\	
| S_z= -3/2 \rangle &= | 1,1,1 \rangle, \qquad   | S_z= -1/2 \rangle = \frac{1}{\sqrt{3}}(| 0,1,1 \rangle + | 1,0,1 \rangle + | 1,1,0 \rangle).\nonumber
\end{align}

\subsection{2D TDVP simulation}

\begin{figure}[h!]
	\centering
	\includegraphics[width=0.8\textwidth]{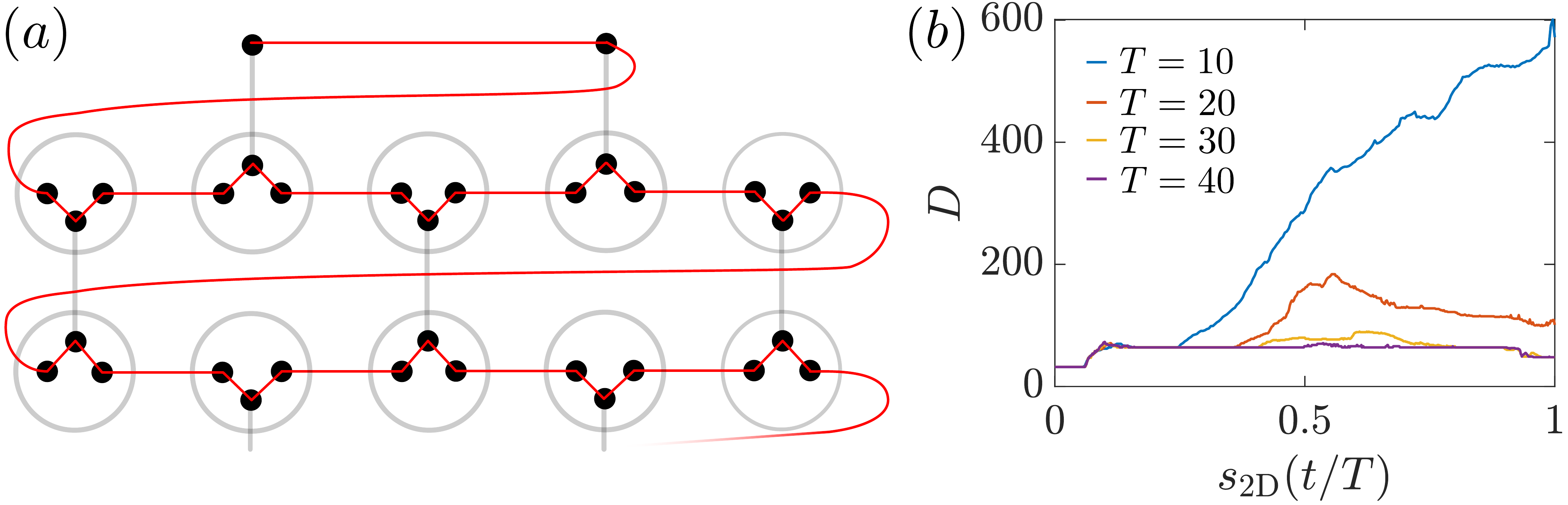}
        \caption{(a)The red curve shows the order of sites of the effective one-dimensional chain for the 2D hexagonal lattice [c.f.Fig.3 in the main text]. (b) The evolution of the bond dimension $D$ for the effective MPS during adiabatic evolution (characterized by $s_{\rm 2D}(t/T)$) for preparing the 2D AKLT of lattice size $5\times 5$ with various evolution time $T$.}
        \label{ak2d_apd_fg}
\end{figure}

We order the 2D hexagonal lattice sites as a one-dimensional chain as shown in \cref{ak2d_apd_fg}(a). In this case, the 2D local Hamiltonian [c.f.~Eq.(3) in the main text] translates to a long-range 1D Hamiltonian, and we use the time-dependent variational principle (TDVP) method~\cite{Haegeman2011,Haegeman2016} to study the time evolution of the system since it potentially has better performance than TEBD method for long-range interacting systems~\cite{paeckel2019time}. 
Also, note that the bond dimension $D$ of this effective MPS grows exponentially with the linear dimension $L$ as $D=O[\exp(L)]$, therefore our numerical method does not scale up to very large system sizes. The adiabatic path with a ground state as $D=2$ PEPS [c.f.~Eq.(6) in the main text] is thus crucial for simulating the adiabatic dynamics of this system up to a system size $N\sim 10 \times 10$ [c.f.~Fig.3 in the main text]. This is demonstrated in \cref{ak2d_apd_fg}(b), that one can see the evolution of the bond dimension $D$ during the adiabatic evolution with various total evolution time $T$. When $T$ is small, the bond dimension increases significantly during the evolution. By increasing $T$, the dynamics become more adiabatic, thus we observe the overall bond dimension gets significantly reduced.

\subsection{Preparation of the 2D AKLT state on the square lattice}

\begin{figure}[h!]

	\centering
	\includegraphics[width=0.75\textwidth]{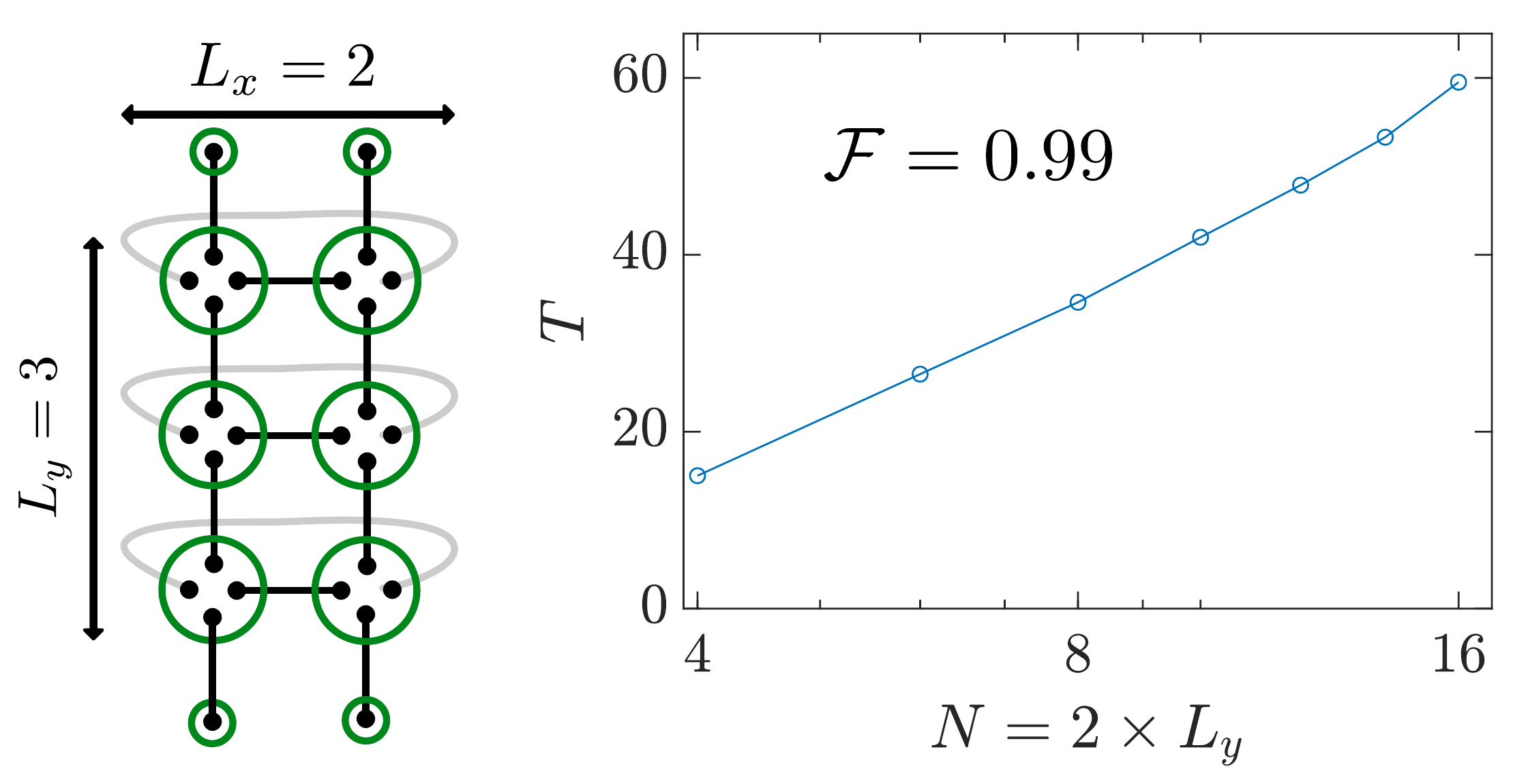}
        \caption{(a) The 2D AKLT state on the square lattice of size $N=2\times 3$, with cylinder boundary condition. Here we use the same notations as that in Fig.1(c) of the main text. (b) The evolution time $T$ needed to prepare the 2D AKLT state on the square lattice with fidelity ${\cal F}=0.99$ for $N=2\times L$. The line is a visual guide.}
        \label{ak_sq_fg}
\end{figure}

To explicitly demonstrate that our method can prepare non-normal high-dimensional PEPS (in contrast to the adiabatic path in Ref.~\cite{Ge2016,cruz2022preparation}), in \cref{ak_sq_fg}(b) we show the evolution time $T$ needed to prepare the AKLT state of spin $S=2$ on the 2D square lattice of size $N=2\times L$ of cylinder boundary condition [c.f.~ \cref{ak_sq_fg}(a)] with fidelity ${\cal F}=0.99$, using our adiabatic path [c.f.~Eq.(6) in the main text] with the interpolation function $s(t,T)_{\rm 2D}\equiv {\sin}^2(\pi t/2T)$. Similar to the case of the hexagonal lattice [c.f.~\cref{apd_ak2d_hex}], we have promoted the virtual qubits in the AKLT state to physical ones. Thus the Hamiltonian during the path [c.f.~Eq.(3) in the main text] are always two-body, where each site contain four spin-1/2 qubits.

The result in \cref{ak_sq_fg}(b) shows that the 2D AKLT state on the square lattice (which is a non-normal PEPS~\cite{Molnar2018}), can be prepared efficiently for system sizes up to $N=2\times 8$. We expect that the scaling in \cref{ak_sq_fg}(b) to persist when we increase the system size further.

\bibliography{library.bib}